\documentclass[a4paper,11pt]{article}
\pdfoutput=1 

\usepackage{jheppub} 

\usepackage[T1]{fontenc} 
\usepackage{multirow}

\newcommand{\ttbar}{t{\bar t}}

\newcommand{\as}{\alpha_s}
\newcommand{\QQbar}{Q{\bar Q}}

\newcommand{\einhalb}{ \frac{1}{2} }

\newcommand{\cA}{\ensuremath{\mathcal{A}}}
\newcommand{\cB}{\ensuremath{\mathcal{B}}}

\newcommand{\cD}{\ensuremath{\mathcal{D}}} 
\newcommand{\cE}{\ensuremath{\mathcal{E}}}
\newcommand{\cF}{\ensuremath{\mathcal{F}}}

\newcommand{\cM}{\ensuremath{\mathcal{M}}}
\newcommand{\cN}{\ensuremath{\mathcal{N}}}
\newcommand{\cO}{\ensuremath{\mathcal{O}}} 

\newcommand{\mss}[1]{ {\mbox{\scriptsize #1}} }

\newcommand{\mycomment}[1]{ } 

\newcommand{\afb}{A_{\rm FB}}

\newcommand{\sia}{\sigma_A}
\newcommand{\sis}{\sigma_S}

\preprint{TTK-16-41, MITP/16-111}

\title{\boldmath Top-quark pair production at next-to-next-to-leading order QCD \\in electron positron collisions}

\author[a]{Long Chen,}
\author[b]{Oliver Dekkers,}
\author[a]{Dennis Heisler,}
\author[a,1]{Werner Bernreuther,\note{Corresponding author.}}
\author[c]{Zong-Guo Si}

\affiliation[a]{Institut f\"ur Theoretische Teilchenphysik und Kosmologie, RWTH Aachen University,\\ 52056 Aachen, Germany}
\affiliation[b]{PRISMA Cluster of Excellence and Institut f\"ur Physik, Johannes-Gutenberg-Universit\"at Mainz, \\ 55099 Mainz, Germany}
\affiliation[c]{School of Physics, Shandong University, Jinan, Shandong 250100, China}

\emailAdd{algeochen@physik.rwth-aachen.de}
\emailAdd{dekkers@uni-mainz.de}
\emailAdd{heisler@physik.rwth-aachen.de}
\emailAdd{breuther@physik.rwth-aachen.de}
\emailAdd{zgsi@sdu.edu.cn}

\abstract{We set up a formalism, within the antenna subtraction framework, for computing the production of a massive quark-antiquark pair 
 in electron positron collisions at  
 next-to-next-to-leading order in the coupling $\alpha_s$ of quantum chromodynamics at the differential level. Our formalism applies to the calculation of any infrared-safe observable.
 We apply this set-up to the production of top-quark top antiquark pairs in the continuum. We compute the production cross section and several distributions.
 We determine, in particular, the top-quark forward-backward asymmetry at order $\alpha_s^2$. Our result agrees with previous computations of this observable. }
 

\begin{document} 
\maketitle
\flushbottom

\section{Introduction}
\label{sec:intro}

The  exploration of the production of top-quark top-antiquark $(\ttbar)$ pairs and their decays is among the 
 core physics issues at future linear or circular electron-positron colliders \cite{AguilarSaavedra:2001rg,Baer:2013cma,Gomez-Ceballos:2013zzn}.
 Simulation studies
  indicate that measurements of the reaction $e^-e^+\to \ttbar$ in the threshold region and in the continuum allow to precisely determine a number of key observables associated with the top quark,
  including its mass, its width, its Yukawa coupling to the 125 GeV Higgs resonance, and its electroweak neutral current couplings
    (cf., for instance, \cite{Moortgat-Picka:2015yla,Vos:2016til} and references therein). Needless to say, 
   precise predictions are required, too, on the theoretical side. A large effort has been made to investigate $\ttbar$ production at threshold. At present the threshold 
    cross section is known at next-to-next-to-next-to-leading order QCD \cite{Beneke:2015kwa}. As far as the production  of $\ttbar$, or more general, the production of a 
     heavy quark-antiquark pair $(Q{\bar Q})$ in the continuum is concerned, differential predictions at next-to-leading order (NLO) QCD have been known for a long time
      for $Q{\bar Q}$ \cite{Jersak:1981sp}   and $Q{\bar Q}$ + jet \cite{Bernreuther:1997jn,Brandenburg:1997pu,Rodrigo:1997gy,Rodrigo:1999qg,Nason:1997tz,Nason:1997nw}
final states. Also the NLO electroweak corrections are known \cite{Beenakker:1991ca,Fleischer:2003kk,Hahn:2003ab,Khiem:2012bp}. 
 Off-shell $\ttbar$ production and decay including non-resonant and interference contributions at 
 NLO QCD was investigated in \cite{Nejad:2016bci}.
 The total $Q{\bar Q}$ cross section $\sigma_{Q{\bar Q}}$  was computed to 
 order $\alpha_s^2$ (NNLO) and order $\alpha_s^3$ in \cite{Gorishnii:1986pz,Chetyrkin:1996cf,Chetyrkin:1997qi,Chetyrkin:1997pn} and \cite{Kiyo:2009gb}, respectively, using 
  approximations as far as the dependence of $\sigma_{Q{\bar Q}}$ on the mass of $Q$ is concerned. (A calculation of $e^-e^+\to \gamma^*\to Q{\bar Q}$ with full
   quark-mass dependence of  $\sigma_{Q{\bar Q}}$ was made in \cite{Dekkers:2014hna}.) A computation of the cross section and of
    differential distributions for $\ttbar$ production at  order $\alpha_s^2$  with full top-mass dependence  was reported in \cite{Gao:2014nva,Gao:2014eea}.

 In this paper we set up a formalism for calculating the electroweak production of a massive quark-antiquark pair,
\begin{equation} \label{eq:QQincl}
 e^-(p_1) e^+(p_2) \to \gamma^*,Z^*(q) \to Q(k_1)~{\bar Q}(k_2) + X  \, ,
\end{equation}
 at order $\alpha_s^2$  and to lowest order in the electroweak couplings within the antenna subtraction framework and apply it to the production of 
  top-quark pairs. Our approach is fully differential and applies to any infrared-finite observable. 
  
  Antenna subtraction is a method for handling infrared (IR) divergences, that is, soft and collinear divergences in higher order QCD calculations
   \cite{Kosower:1997zr,Kosower:2003bh,GehrmannDeRidder:2005cm,Currie:2013vh}.
  The general features of the method at NNLO QCD were developed in ref.~\cite{GehrmannDeRidder:2005cm}.
   For QCD processes with massive quarks the antenna subtraction terms at NLO were 
   determined in refs.~\cite{GehrmannDeRidder:2009fz,Abelof:2011jv}. As to applications of this method to hadronic $\ttbar$ production, partial results were obtained 
   in refs.~\cite{Abelof:2011ap,Abelof:2012he,Abelof:2014fza,Abelof:2014jna,Abelof:2015lna}. For the computation of the reaction  \eqref{eq:QQincl} at the differential level
    we use the unintegrated and integrated NNLO real radiation antenna subtraction terms and the NNLO real-virtual antenna functions 
    worked out in \cite{Bernreuther:2011jt,Bernreuther:2013uma} and  \cite{Dekkers:2014hna}, respectively. 
    We recall that alternative methods for handling IR divergences have been successfully applied to NNLO QCD processes involving top quarks. 
    The method of \cite{Czakon:2010td,Czakon:2011ve} was used in the computation of the hadronic $\ttbar$ production cross section \cite{Czakon:2013goa} and of 
    differential distributions \cite{Czakon:2015owf}. The results obtained by \cite{Gao:2014nva,Gao:2014eea} for \eqref{eq:QQincl} 
     are based on a NNLO generalization of a phase-space 
     slicing method \cite{Gao:2012ja,vonManteuffel:2014mva}.
             
       This paper is organized as follows. We recapitulate in the next section the calculation of the differential cross section of  \eqref{eq:QQincl}  at order 
       $\as$ using the  antenna subtraction method.  Section~\ref{sec:nnlo} contains a detailed exposition of how to compute within this framework the differential cross section and 
       distributions of IR-safe observables at  order $\as^2$.  In section~\ref{sec:restt} we apply this formalism to top-quark pair production above the $\ttbar$ threshold. We compute 
       the total $\ttbar$ cross section, a number of differential distributions, and the top-quark forward-backward asymmetry at order $\as^2$. We compare also with existing results.
       We conclude in section~\ref{sec:sumconc}. Appendix~\ref{sec:AppA} contains details about the momentum mappings in the three- and four-particle phase spaces 
        with massive quarks that are required for the antenna subtraction terms at NLO and NNLO QCD.

 \section{The differential cross section at LO and NLO QCD} 
 \label{sec:lonlo}
 For completeness and for setting up our notation we outline in this section the computation of the differential cross
 section for $ e^-e^+\to Q {\bar Q} X$ at order $\alpha_s$ within the antenna subtraction method.
 Here $Q$ denotes a massive quark, for instance, the  $b$ or $t$ quark.
  We work in QCD with $n_f$ massless quarks $q$ and one massive quark $Q$.
  All matrix elements in this and in the 
  following section~\ref{sec:nnlo} refer to renormalized matrix elements. 
   We define the mass of $Q$, denoted by $m_Q$, in 
  the on-shell scheme while the QCD coupling $\alpha_s$ is defined in the $\overline{\rm MS}$ scheme.
   Dimensional regularization is used to handle IR singularities that appear in intermediate steps of
   our calculation.

\subsection{LO QCD} 
\label{susec:lo}

To zeroth order in $\alpha_s$ we consider
\begin{equation} \label{eeQQLO}
 e^-(p_1)~e^+(p_2) \to \gamma^*,Z^*(q) \to Q(k_1)~{\bar Q}(k_2) \, ,
\end{equation}
 where  $Q$ denotes a massive quark.
The corresponding leading-order (LO) (differential) cross section for unpolarized
$e^-e^+$ collisions is given by
\begin{equation}
\int \, d\sigma_{\mss{LO}} = \frac{1}{8 s} \, N_c 
\int \, d \Phi_2 \,  
 \cF_2^{(2)}( k_1, k_2 ) \,
\big| \cM_2^0 ( 1_Q, 2_{\bar{Q}} ) \big|^2 \,,
\label{sigma::LO}
\end{equation}
where the color-stripped two-parton Born amplitude $\cM_2^0 ( 1_Q,
2_{\bar{Q}} )$ is defined in eq.~\eqref{eq::M02_S2QQ} below. 
We use here and below, as  ref.~\cite{GehrmannDeRidder:2005cm}, symbolic labels $i_X$ in order to 
 display the type $X$ and the four-momentum $k_i$ of a final-state parton in the matrix elements. For instance, 
  $1_Q$ denotes a massive quark with momentum $k_1$.
 Here and in the following,
 summation over the spins and colors of the partons in the final state is implicit.
The factor $1/8s$
 is the product of the spin averaging factor for the initial state and the flux
factor, the variable $s=(p_1+p_2)^2$, and $N_c=3$. 
In general the jet function or measurement function, which must be infrared-safe,
 is denoted by  $\cF_n^{(m)}(k_j)$. It refers to a $n$-jet observable constructed 
out of a pair of $Q,{\bar Q}$ and $(m-2)$ massless partons in the final state.
Here and in the following section we put for definiteness $n=2$, but we emphasize
 that our analysis applies to any infrared-safe observable.
 The  $m$-particle phase space measure $d \Phi_m$ in $D$ space-time dimensions is
\begin{equation} \label{eq:phaseD}
d \Phi_m ( k_j ; q ) = (\mu^{4-D})^{m-1}\prod^{m}_{i = 1}
 \!\frac{d^{D-1} {k_i}}{ (2\pi)^{D-1} 2k_i^0} 
\left( 2 \pi \right)^{D}
\delta^{(D)}\! \left( q - \sum^m_{i = 1} k_i  \right) \, ,
\end{equation}
 where $\mu$ is a mass scale. 
%

%
 
\subsection{NLO QCD} 
\label{susec:nlo}
The computation of the NLO QCD correction $d\sigma_1$ to \eqref{sigma::LO} involves the interference of the
Born and one-loop amplitude of \eqref{eeQQLO} and the squared Born amplitude 
 of the three-parton final state
\begin{equation} \label{RQQg}
 e^-(p_1)~e^+(p_2) \to \gamma^*,Z^*(q) \to Q(k_1)~{\bar Q}(k_2) + g(k_3)  \, .
\end{equation}
 We recall that in a subtraction scheme for handling the IR divergences, the 
NLO correction to the LO cross section or to a differential distribution can be written
as follows:
\begin{equation} 
  \int\, d\sigma_{1} = \int_{\Phi_3} \! \Big[
    \left(d\sigma_{Q\bar{Q}g}^{R}\right)_{\epsilon=0} 
    - \left(d\sigma_{Q\bar{Q}g}^{S}\right)_{\epsilon=0}
  \Big]
  + \int_{\Phi_2}  \left[ \,
    d\sigma_{Q\bar{Q}}^{V} + \int_1 d\sigma_{Q\bar{Q}g}^{S}
  \right]_{\epsilon=0} \, ,
  \label{eq:subtrNLO}
\end{equation}
where $\epsilon=(4-D)/2$ is the parameter of dimensional regularization and the subscripts $\Phi_n$ denote 
 $n$-particle phase-space integrals. The second term in the first and second square bracket of 
 \eqref{eq:subtrNLO} is the unintegrated and integrated subtraction term that renders the difference, respectively sum of 
 the terms in the square brackets finite in $D=4$ dimensions. Throughout this paper,
  the symbol
 $\int_n$ indicates the analytic integration over the phase space of $n$ unresolved partons in $D\neq 4$ dimensions.
 Within the antenna  framework, the NLO subtraction terms 
 required in \eqref{eq:subtrNLO} were computed in \cite{GehrmannDeRidder:2009fz}.

 The NLO real and virtual corrections to the LO differential cross section,
$d\sigma_{Q\bar{Q}g}^{R}$ and $d\sigma_{Q\bar{Q}}^{V}$, are given by 
\begin{eqnarray}
d\sigma_{Q\bar{Q}g}^{R} & = & \frac{1}{8 s}
\left( 4 \pi \alpha_s \right)
\left( N_c^2 - 1 \right) 
d \Phi_3\,  
\cF_2^{(3)}( k_1, k_2, k_3 ) \,
\big| \cM_3^0 ( 1_Q, 3_g, 2_{\bar{Q}} ) \big|^2  \,,
\label{dsigma::NLO::R}
\\[1.5ex]
d\sigma_{Q\bar{Q}}^{V} & = & \frac{1}{8 s}
\left( \frac{ \alpha_s }{ 2 \pi } \right) \bar{C}( \epsilon ) 
\left( N_c^2 - 1 \right) 
d \Phi_2 \,  
\cF_2^{(2)} ( k_1, k_2 ) \,
\delta \cM_2^1 ( 1_Q, 2_{\bar{Q}} )  \, , 
\label{dsigma::NLO::V}
\end{eqnarray}
where
\begin{equation} \label{eq:defCeps}
 \bar{C}(\epsilon) = 8 \pi^2 C( \epsilon ) = ( 4 \pi )^\epsilon 
e^{-\epsilon \gamma_E } 
\end{equation}
and $\gamma_E = 0.57721 \ldots $ denotes the Euler--Mascheroni constant.
As before summation over the spins and colors is understood.
 We use the following shorthand notation for the interference of the tree-level and one-loop
two-parton amplitude:
\begin{equation}
\delta \cM_2^1  \! \left( i_Q, j_{\bar{Q}} \right) 
= 2 \mbox{Re} \left[ \cM^{0*}_2  \! \left(  i_Q, j_{\bar{Q}} \right) 
\cM_2^1  \! \left(  i_Q, j_{\bar{Q}} \right) \right] \, .
\label{eq:delLONLO2}
\end{equation} 
In the formulas 
\eqref{sigma::LO}, \eqref{dsigma::NLO::R}, and \eqref{eq:delLONLO2} we
have introduced  color stripped partial amplitudes
$\cM_2^0$, $\cM_2^1$, and $\cM_3^0$ where  QCD coupling factors are taken out,
but electroweak couplings are included.
 These quantities
are related to the tree-level and the renormalized one-loop matrix elements of
$e^- e^+ \to Q \bar{Q}$
and the tree-level matrix element of $e^- e^+\to Q \bar{Q} g$.
For reference in the next section we give here the expansion to NNLO QCD of the matrix elements
 of these processes:
\begin{eqnarray}
    M ( e^- e^+ \to Q \, \bar{Q}) 
    & = & 
    \delta_{i_1 i_2} \, 
    \cM_2^0 \! \left( 1_Q , 2_{\bar{Q}} \right) 
    \nonumber \\
    & + & 
     \left( \frac{ \alpha_s  }{ 2 \pi } \right)   
      \bar{C}( \epsilon )\,
      2 C_F \,
      \delta_{i_1 i_2} \, 
      \cM_2^1 \! \left( 1_Q, 2_{\bar{Q}} \right)   \nonumber \\
    & + & \left( \frac{ \alpha_s  }{ 2 \pi } \right)^2   
      \left( \bar{C}( \epsilon ) \right)^2
      \delta_{i_1 i_2} \, 
      \cM_2^2 \! \left( 1_Q, 2_{\bar{Q}} \right) 
      + \cO \! \left( \alpha_s^3 \right) , 
    \label{eq::M02_S2QQ}
 \end{eqnarray}   
 \begin{eqnarray}
	M (e^- e^+\to Q \bar{Q} g )
	& = &
	g_s\,
	\sqrt{2} \, T^{a_3}_{i_1 i_2} \, \bigg\{
	\cM_3^0\! \left( 1_Q, 3_g, 2_{\bar{Q}} \right) 
	+ \left( \frac{ \alpha_s }{2 \pi} \right) 
	\bar{C}(\epsilon)
	\nonumber \\
	& & {} \times \left[
	N_c \,
	\cM^{\mss{1,lc}}_3 \! \left( 1_Q, 3_g, 2_{\bar{Q}} \right) 
	- \frac{1}{N_c} \, 
	\cM^{\mss{1,sc}}_3 \! \left( 1_Q, 3_g, 2_{\bar{Q}} \right) 
	\right.
	\nonumber \\
	& & {} + n_f
	\cM^{1,f}_3 \! \left( 1_Q, 3_g, 2_{\bar{Q}} \right) 
	+
	\cM^{1,F}_3 \! \left( 1_Q, 3_g, 2_{\bar{Q}} \right) 
	+
	\cM^{1,{\rm tr}}_3 \! \left( 1_Q, 3_g, 2_{\bar{Q}} \right) 	
	\bigg] \nonumber \\
	& & {} + \cO \! \left( \alpha_s^2 \right) 
	\bigg\} \, ,
	\label{eq::M03_S2QQgA}
\end{eqnarray}
where $i_1$ ($i_2$) denotes the color index of the heavy quark
(antiquark), $a_3$ is the color index of the gluon, $g_s=\sqrt{4\pi\alpha_s}$,
 $C_F=(N_c^2-1)/(2N_c)$, and the 
generators of SU(3)$_c$ are normalized according to ${\rm tr}(T^aT^b)=T_R \delta_{ab}$ with $T_R=1/2$.
The number of massless quarks is denoted by $n_f$.
The renormalized two-loop two-parton amplitude $\cM_2^2$, which can be decomposed into different color structures, and  
the renormalized  1-loop three-parton amplitude in the square bracket of \eqref{eq::M03_S2QQgA} are required in the
 next section. The labels `lc' and `sc' in \eqref{eq::M03_S2QQgA} refer to leading and subleading color, respectively.
 The terms 
$\cM^{1,f}_3$ and $\cM^{1,F}_3$ are the contributions from the massless and massive quark loop, respectively,
 that enter via the wave-function renormalization of the external gluon. The term $\cM^{1,{\rm tr}}_3$ denotes the quark triangle contributions,
 where the axial current couples to quark triangles, summed over all quark flavors $(u_i,d_i)$,
 which disintegrate into a real and virtual gluon that splits into $Q{\bar Q}$. This term, which is ultraviolet- and infrared-finite, involves weak couplings 
 of $q\neq Q$ and constitutes in this sense a non-universal correction to the leading-order $Q{\bar Q}$ cross section.

Let's proceed with the discussion of the NLO cross section.
The squared Born matrix element $\big| \cM_3^0 \big|^2$ of the real radiation
correction \eqref{dsigma::NLO::R} diverges when the gluon momentum $k_3$
becomes soft. Within the antenna method this singularity is regularized by
 constructing a subtraction term that coincides with \eqref{dsigma::NLO::R}
  in this singular limit. 
 The subtraction term and its integrated
form (integrated over the phase space of the unresolved gluon) are:
\begin{eqnarray}
  d\sigma_{Q\bar{Q}g}^S & = & \frac{1}{8 s}
  \left( 4 \pi \alpha_s \right)
  \left( N_c^2 - 1 \right) 
  d \Phi_3 \!\left( k_1,k_2, k_3; q \right)  
  \cF_2^{(2)} \! \left( \widetilde{ k_{13} }, \widetilde{k_{32}} \right) \,
  \nonumber 
  \\
  & & {} \times 
  A^0_3\! \left( 1_Q, 3_g, 2_{\bar{Q}} \right)
  \Big| \, \cM_2^0 
  \! \left( \widetilde{(13)}_Q, \widetilde{(32)}_{\bar{Q}} \right) \!
  \Big|^2 \,,
  \label{sub::NLO}
  \\[1.5ex]
  \int_1 d\sigma_{Q\bar{Q}g}^S & = & \frac{1}{8 s}
  \left( \frac{ \alpha_s }{ 2 \pi} \right)
  \bar{C}( \epsilon )\,
  \left( N_c^2 - 1 \right) 
  d \Phi_2 ( k_1,k_2 ; q )\,  
  \cF_2^{(2)} (  k_{1}, k_{2} ) \,
  \nonumber 
  \\
  & & {} \times 
  \cA^0_3 \! \left( \epsilon, \mu^2 /s ; y \right) 
  \big| \, 
  \cM_2^0 \! \left( 1_Q, 2_{\bar{Q}} \right) \!
  \big|^2  \,  ,
\end{eqnarray}
where $\mu$ is the mass parameter of dimensional regularization, and 
\begin{equation} \label{eq:defybet}
y=\frac{1-\beta}{1+\beta} \, , \quad \beta=\sqrt{1-4m_Q^2/s} \, .  
\end{equation}
The three-parton tree-level massive quark-antiquark antenna function $A^0_3$
and its integrated counterpart $\cA^0_3$ were derived in \cite{GehrmannDeRidder:2009fz,Abelof:2011jv}. 
The integrated antenna function $\cA^0_3$ contains an explicit IR pole $\propto 1/\epsilon$ that cancels 
 the corresponding IR pole in $d\sigma_{Q\bar{Q}}^{V}$.  
In \eqref{sub::NLO} the
matrix element $\cM_2^0$ and the measurement function $\cF_2^{(2)}$
are evaluated with
redefined on-shell momenta $\widetilde{ k_{13} }$, $\widetilde{k_{32}}$ that are
obtained from ${k_1,k_2,k_3}$ by an appropriate phase-space mapping \cite{Abelof:2011ap}. 
A method to construct  $\widetilde{ k_{13} }$, $\widetilde{k_{32}}$ is given in appendix~\ref{suse:map3p}.

\section{The differential cross section at NNLO QCD}
\label{sec:nnlo}

The second-order term $d\sigma_{2}$ in the expansion in powers of $\alpha_s$ 
of the differential cross section of  \eqref{eq:QQincl}, $d\sigma=d\sigma_{\rm LO}+
d\sigma_{1} + d\sigma_{2} +{\cO}(\as^3)$,  receives the following contributions:
i) the double virtual correction $d\sigma^{VV}_{\rm NNLO}$  associated
with the  matrix element of 
$e^-e^+ \to \QQbar$ to order $\alpha_s^2$
(i.e., 2-loop times Born and 1-loop squared), 
 ii) the real-virtual cross section
 $d\sigma^{RV}_{\rm NNLO}$ associated with the matrix element of 
$e^-e^+  \to \QQbar g$ to order   $\alpha_s^2$     (1-loop times Born),
iii) the double real contribution $d\sigma^{RR}_{\rm NNLO}$
associated with  the squared Born amplitudes  
$ e^-e^+ \to \QQbar gg$, 
$ e^-e^+ \to \QQbar q{\bar q}$ 
(where   $q$ denotes a massless quark).
 Above the $4Q$ threshold,  
 $e^-e^+ \to \QQbar\QQbar$ contributes, too.
 The latter contribution is IR finite and is of no concern for the purpose of
this section. 

Apart from the $\QQbar\QQbar$   contribution,
the terms i), ii), iii) are IR divergent. 
Within the  subtraction method the second order correction $d\sigma_{2}$, where the different pieces 
are separately finite, 
 is constructed schematically as follows:
\begin{eqnarray}
\int d\sigma_{2} = & \int_{\Phi_4}\left(d\sigma^{RR}_{\rm NNLO}  - d\sigma^{S}_{\rm NNLO}\right)
+  \int_{\Phi_3}\left(d\sigma^{RV}_{\rm NNLO}  -d\sigma^{T}_{\rm NNLO}\right) \nonumber \\
&  + \left( \int_{\Phi_2} d\sigma^{VV}_{\rm NNLO} +  \int_{\Phi_3} d\sigma^{T}_{\rm NNLO} 
+ \int_{\Phi_4} d\sigma^{S}_{\rm NNLO} \right) \, .
\label{eq:sub2NNLO}
\end{eqnarray}
The integrands $d\sigma^{S}_{\rm NNLO}$ and $d\sigma^{T}_{\rm NNLO}$ denote 
the double-real subtraction terms (for $\QQbar q{\bar q}$ and  $\QQbar gg$) 
and the real-virtual subtraction term, respectively. 
We discuss in turn the various terms in \eqref{eq:sub2NNLO} in some detail.

\subsection{Double real-radiation corrections}
\label{suse:RR}

 In this subsection we discuss how to compute the first term on the right-hand side of the
  first line in \eqref{eq:sub2NNLO} with the antenna subtraction method.

\subsubsection*{The $Q{\bar Q}q{\bar q}$ final state:} 

First we consider the reaction
\begin{equation} \label{RQQqq}
 e^-(p_1) e^+(p_2) \to \gamma^*,Z^*(q) \to Q(k_1)~{\bar Q}(k_2) + q(k_3)~{\bar q}(k_4)  \, ,
\end{equation}
where $q$ denotes a massless quark.
 The corresponding tree-level amplitude, decomposed into color-stripped 
 subamplitudes with the QCD coupling factored out,
 is given by
\begin{eqnarray} 
 M\! \left(e^-e^+ \to Q\bar{Q} q \bar{q} \right) & = & 
 \left( 4 \pi \alpha_s \right)
\left(
\delta_{i_1 i_4} \delta_{i_3 i_2} 
- \frac{1}{N_c} \, \delta_{i_1 i_2} \delta_{i_3 i_4} \right) 
\bigg( \cM^{Q}_4 + \cM^{q}_4  \bigg) \, .
\label{QQqqAmp}
\end{eqnarray}
The color indices of the quarks and antiquarks are labeled by
$i_1, \ldots, i_4$. The matrix element $\cM^{Q}_4$
($\cM^{q}_4$) corresponds to the
subamplitude where the massless (massive) quark-antiquark pair is produced by
the splitting of the virtual gluon radiated off one of the quarks produced by the
virtual photon or $Z$ boson.  The matrix element
\eqref{QQqqAmp} yields the unsubtracted differential cross section summed over colors and summed/averaged over 
all spins:
\begin{eqnarray}
\sum_q d \sigma^{Q \bar{Q} q \bar{q} }_\mss{NNLO} & = & 
\cN d \Phi_4\,  \cF^{(4)}_2 ( k_1, k_2, k_3, k_4 ) \nonumber 
  \\[-1ex]
  & & {} \times
  \bigg\{ \, n_f \,
 \big| \cM^{Q}_4\big|^2 
  + \sum_q   \big| \cM^{q}_4 \big|^2 
+ \sum_q  2 \,\mbox{Re}\left[\left(\cM^{Q}_4\right)^\ast  \cM^{q}_4  \right] \bigg\},
  \label{eq:sigmaQQqq}
\end{eqnarray}
where the sum is over all $n_f$ massless quark flavors and
\begin{equation}\label{eq:defN}
 \cN =
\frac{1}{8 s} \left( 4 \pi  \alpha_s \right)^{2} 
\left( N_c^2 - 1 \right) \, .
\end{equation}
 The second and third term in the second line of eq.~\eqref{eq:sigmaQQqq} contain the 
 electroweak couplings of the massless quarks $q$. Thus these terms are non-universal
 QCD corrections to the leading-order differential $Q{\bar Q}$ cross section.
  Moreover, these two terms do not become infrared singular in the four-parton phase space.
  Only the first term in the second line of eq.~\eqref{eq:sigmaQQqq} requires subtraction.

   We define a subtracted differential cross section by subtracting three terms from \eqref{eq:sigmaQQqq}.
\begin{equation}  \label{eq:QQqq-sub}
d \sigma^{{\rm sub},Q \bar{Q} q \bar{q}}_\mss{NNLO} = 
\sum\limits_q \left( d \sigma^{Q \bar{Q} q \bar{q}}_\mss{NNLO} - d \sigma^{S,a,Q \bar{Q} q \bar{q}}_\mss{NNLO} 
- d\sigma^{S,b,2, Q \bar{Q} q \bar{q}}_\mss{NNLO} - \sigma^{S,b,1, Q \bar{Q} q \bar{q}}_\mss{NNLO} \right).
\end{equation}
By construction the sum of the three subtraction terms is such that it coincides with \eqref{eq:sigmaQQqq}
 in all single and double unresolved limits, i.e., when the massless quarks become collinear and/or soft.
 Thus, \eqref{eq:QQqq-sub} is free of IR divergences and can be integrated over the four-parton phase space
numerically in $D=4$ dimensions. \\
The term $d \sigma^{S,a,Q \bar{Q} q \bar{q}}_\mss{NNLO}$ subtracts the singularities associated with the single 
unresolved configurations from  the first term in the curly bracket of \eqref{eq:sigmaQQqq}.
 Within the antenna method, it is given by
\begin{align} \label{eq:SQQqqa}
d \sigma^{S,a,Q \bar{Q} q \bar{q}}_\mss{NNLO} & =  {\cal N} d \Phi_4 ( k_1, k_2, k_3, k_4; q )
\nonumber \\
 & {} \times \bigg[ \, \frac{1}{2} E^0_3\! \left( 1_Q, 3_q, 4_{\bar{q}} \right) 
 \left|\cM_3^0\left( \widetilde{ (13) }_Q, \widetilde{ (34) }_g,
2_{\bar{Q}}\right)\right|^2 \cF^{(3)}_2 \!  \left( \widetilde{ k_{13} },
\widetilde{ k_{34} }, k_2 \right) 
\nonumber \\
 & + \frac{1}{2} E^0_3\! \left(2_{\bar{Q}}, 3_q, 4_{\bar{q}} \right)
  \left|\cM_3^0 \left( 1_Q , \widetilde{ (34) }_g, \widetilde{
(42)}_{\bar{Q}} \right)\right|^2 \cF^{(3)}_2 \! \left(  k_1, \widetilde{ k_{34}
}, \widetilde{ k_{42} } \right) \bigg]. 
\end{align}
 The quark-gluon antenna $E^0_3$ with a massive radiator (anti)quark is 
  given in \cite{GehrmannDeRidder:2009fz}. The color-stripped tree-level $Q{\bar Q}g$ matrix element 
  $\cM_3^0$ is defined in \eqref{eq::M03_S2QQgA}.
 The momenta $\widetilde{ k_{ij} }$ and $\widetilde{ k_{jk}}$ are redefined on-shell momenta, 
 constructed from linear combinations of the
momenta $k_i$, $k_j$ and $k_k$ \cite{Abelof:2011ap,AbelGerh2016}. The $3\to 2$ mappings must be such that
 they define remapped on-shell momenta that have the correct soft and collinear limits. In appendix~\ref{sec:AppA}
 we discuss the $3\to 2$ and $4\to2$ mappings that are required for the double real-radiation subtraction terms. 
 A numerical method to construct the mapped momenta that appear in  \eqref{eq:SQQqqa} is given in appendix~\ref{suse:map3p}.

The subtraction term for removing the singularities of  \eqref{eq:sigmaQQqq} due to  the double unresolved configuration,
where both $q$ and ${\bar q}$ become soft, is
\begin{align}
d \sigma^{S,b,2, Q \bar{Q} q \bar{q}}_\mss{NNLO} & = &  {\cal N}  d \Phi_4 
\; B^0_4 ( 1_Q, 3_q, 4_{\bar{q}}, 2_{\bar Q} )  \left|
\cM_2^0\left(\widetilde{(134)}_Q, \widetilde{(342)}_{\bar Q}\right)\right|^2
\cF_2^{(2)}\left(\widetilde{ k_{134}},\widetilde{ k_{342}}\right) \, .
\label{eq:Sb2qqbar}
\end{align}
The antenna function $B^0_4$ is given in \cite{Bernreuther:2011jt} and
 the tree-level matrix element $\cM_2^0$ is defined in  \eqref{eq::M02_S2QQ}.
 The momenta $\widetilde{ k_{ikl} }$ and $\widetilde{ k_{jkl} }$ are linear
combinations of the momenta $k_i$, $k_j$, $k_k$, $k_l$ obtained from a $4\to 2$ mapping, cf. appendix~\ref{suse:map4p}. \\
The  function $B^0_4$ develops singularities in 
the single unresolved limits that are subtracted by the additional term
\begin{flalign}
d \sigma^{S,b,1, Q \bar{Q} q \bar{q}}_\mss{NNLO}  =     - \frac{{\cal N}}{2}  d \Phi_4 ( k_1, k_2, k_3, k_4; q ) \nonumber & \\[0.1cm]
 \times \bigg[ E^0_3\! \left(
1_Q, 3_q, 4_{\bar{q}} \right) A^0_3 \!
\left( \widetilde{(13)}_Q, \widetilde{(34)}_g, 2_{\bar{Q}} \right) 
&\left|\cM_2^0\left(\widetilde{(\widetilde{(13)}\widetilde{(34)})_Q},\widetilde{(\widetilde{(34)}2)_{\bar Q}}\right)\right|^2
\cF_2^{(2)}\left(\widetilde{k_{\widetilde{(13)}\widetilde{(34)}}}, \widetilde{k_{\widetilde{(34)}2}}\right) \bigg. 
\nonumber \\[0.1cm]
 \bigg.  + E^0_3\! \left(2_{\bar{Q}}, 4_{\bar{q}}, 3_q \right) A^0_3 \!
\left( 1_Q, \widetilde{(34)}_g, \widetilde{(42)}_{\bar{Q}} \right) 
&\left|\cM_2^0\left(\widetilde{(1\widetilde{(34)})_Q}, \widetilde{(\widetilde{(34)}\widetilde{(42)})_{\bar Q}}\right)\right|^2
\cF_2^{(2)}\left(\widetilde{k_{1\widetilde{(34)}}}, \widetilde{k_{\widetilde{(34)}\widetilde{(42)}}}\right) \bigg] \, .
\label{eq:Sb1qqbar}
\end{flalign}
The arguments of the antenna functions $A^0_3$ are mapped momenta obtained by the $3\to 2$ mappings described in appendix~\ref{suse:map3p}.  
The arguments of the Born matrix elements and of the measurement functions in \eqref{eq:Sb1qqbar} are obtained by two consecutive $3\to 2$
 mappings described in eqs.~\eqref{eq:it1} and  \eqref{eq:it2} of appendix~\ref{suse:map4p}.
 These two iterated $3\to 2$ mappings are necessary for being able to perform the integration over the antenna phase space of the 
  unresolved parton in analytic fashion and obtain the integrated subtraction term defined in \eqref{eq:TcQQg} below.
  
  Eq.~\eqref{eq:QQqq-sub} is not yet the appropriate expression for numerical evaluation. 
   In the antenna framework there is a subtlety associated with angular correlations \cite{Weinzierl:2006ij,GehrmannDeRidder:2007jk,Glover:2010im,Abelof:2011ap}.
   The gluon radiated off a $Q$ or $\bar Q$ that splits into $q{\bar q}$ leads to angular correlations in the unsubtracted squared matrix element. 
   However, the type-a subtraction term \eqref{eq:SQQqqa} that was constructed to take care of the single unresolved limit of the squared matrix element when the $q$ and $\bar q$
    become collinear, is composed of products of spin-averaged three parton antenna functions and three-parton Born matrix elements. That is, the type-a subtraction term does not 
    contain these angular terms and, therefore, does not have the same local singular behaviour as  the unsubtracted squared matrix element. 
    The four-parton antenna function $B^0_4$ and thus the subtraction term \eqref{eq:Sb2qqbar}, which takes care of the double unresolved limit, 
     contains these angular correlation terms, but the subtraction term \eqref{eq:Sb1qqbar}, which ensures that the complete subtraction term has no singularities in the 
     single unresolved region, does not. However, these angular correlations in the unsubtracted squared matrix element and in the subtraction term \eqref{eq:Sb2qqbar} are averaged 
     out after integration over the azimuthal angle $\phi$ between the spatial parts of the light-like momenta $k_3, k_4$ and the collinear direction $k=k_3+k_4$.
     It can be shown \cite{Weinzierl:2006ij} that the functional dependence on $\phi$ of the squared matrix element in the collinear limit is proportional to $\cos(2\phi+\alpha)$.
     This suggests \cite{Weinzierl:2006ij,GehrmannDeRidder:2007jk,Glover:2010im,Abelof:2011ap} 
     that the angular correlations can be averaged out by combining, for each final-state momentum configuration,
     two points in phase space with azimuthal angles $\phi$ and $\phi+\pi/2$.
     Thus we evaluate  $\big| \cM^{Q}_4\big|^2$ in  \eqref{eq:sigmaQQqq} and  the subtraction term \eqref{eq:Sb2qqbar} for each set of momenta $k_1,k_2,k_3,k_4$
      also for  $k_1, k_2, k_{3r}, k_{4r}$ and take the average. The 4-momenta $k_{3r}, k_{4r}$ are obtained by rotating the spatial parts of $k_3,k_4$ by an angle $\pi/2$
       around the collinear axis ${\bf k}={\bf k_3}+{\bf k_4}$. By sampling the phase space in regions where $k_3\cdot k_4/s \lesssim 10^{-8}$ we checked that this 
       procedure provides a subtraction term that is a very good approximation to the squared matrix element in the single and double unresolved limits.

\subsubsection*{The  $Q \bar{Q} g g$ final state:} 

Next, we consider the reaction
\begin{equation} \label{reac1}
 e^-(p_1) e^+(p_2) \to \gamma^*,Z^*(q) \to Q(k_1) {\bar Q}(k_2) + g(k_3) g(k_4)  \, .
\end{equation}
The corresponding 
tree-level matrix element can be decomposed into color-ordered subamplitudes as follows: 
\begin{eqnarray}
M (e^-e^+ \to Q \bar{Q} g g ) & = &
\left( 4 \pi \alpha_s \right) 2
\left[ 
  \left( T^{a_3} T^{a_4} \right)_{i_1 i_2}
  \cM^g_4 ( 1_Q, 3_g, 4_g, 2_{\bar{Q}} ) 
\right.
\nonumber 
\\
& & \left. {}
+ \left( T^{a_4} T^{a_3} \right)_{i_1 i_2} 
  \cM^g_4 ( 1_Q, 4_g, 3_g, 2_{\bar{Q}} )
\right]  \, .
\label{eq:MQQgg}
\end{eqnarray}
The unsubtracted differential cross section, summed over all colors 
and summed/averaged over all spins, is given by
\begin{eqnarray}
d \sigma^{Q \bar{Q} g g}_\mss{NNLO} & = & \einhalb\, {\cN} d \Phi_4 ( k_1,k_2, k_3, k_4; q )\, 
 {\cF}^{(4)}_2( k_1, k_2, k_3, k_4 )
\nonumber \\
& &  \times 
 \left[ N_c \big( |\cM^g_4 ( 1_Q, 3_g, 4_g, 2_{\bar{Q}} )|^2
+ |\cM^g_4 ( 1_Q, 4_g, 3_g, 2_{\bar{Q}} )|^2 \big)  
- \frac{1}{N_c}\, {\cM}_{\rm sc}
 \right] \, , \qquad
\label{eq:unsigQQgg}
\end{eqnarray}
 where the subleading color term is 
 \begin{equation}\label{eq:M4sub}
{\cM}_{\rm sc} = \big| \cM^g_4 ( 1_Q, 3_g, 4_g, 2_{\bar{Q}} ) 
+ 	\cM^g_4 ( 1_Q, 4_g, 3_g, 2_{\bar{Q}} ) \big|^2  \, .
\end{equation}
 The factor $1/2$ in \eqref{eq:unsigQQgg} is due to Bose symmetry and the factor ${\cN}$ is 
 defined in \eqref{eq:defN}.

 In the subleading color term  ${\cM}_{\rm sc}$ both gluons are 
  photon-like, i.e., no non-abelian gluon vertices are involved.
Hence, when the two  gluons become collinear, this term does not become singular.

In analogy to \eqref{eq:QQqq-sub} we define a subtracted
   differential cross section by subtracting three terms from  \eqref{eq:unsigQQgg}:
\begin{equation}  \label{eq:QQgg-sub}
d \sigma^{{\rm sub},Q \bar{Q} gg}_\mss{NNLO} = 
 d \sigma^{Q \bar{Q} gg}_\mss{NNLO} - d \sigma^{S,a,Q \bar{Q} gg}_\mss{NNLO} 
 - d \sigma^{S,b,2,Q \bar{Q} gg}_\mss{NNLO}  - d \sigma^{S,b,1,Q \bar{Q} gg}_\mss{NNLO}  \, .
\end{equation}
It is by construction free of IR divergences and can be integrated over the four-parton phase space
numerically in $D=4$ dimensions. 
As in the $Q{\bar Q}q{\bar q}$ case  $d\sigma_\mss{NNLO}^{S,\, a\; Q\bar{Q}gg}$ and 
 $d\sigma_\mss{NNLO}^{S,\, b, 2\; Q\bar{Q}gg}$ cover the singularities  of \eqref{eq:unsigQQgg} due to single-unresolved and 
double-unresolved configurations, respectively, that is, when the gluons become collinear and/or soft. The term
 $d \sigma^{S,b,1,Q \bar{Q} gg}_\mss{NNLO}$ subtracts the singularities of 
 $d\sigma_\mss{NNLO}^{S,\, b, 2\; Q\bar{Q}gg}$ in the single unresolved limits.
 
Within the antenna method, these three subtraction terms are given by 
\begin{eqnarray}\label{sub:gga}
\lefteqn{d \sigma^{S,a, Q \bar{Q} g g}_\mss{NNLO} = \einhalb\, {\cN} \, 
d \Phi_4 ( k_1, k_2, k_3, k_4; q ) } \quad
\nonumber \\[1ex]
& & {} \times \bigg\{ \, N_c \bigg[  d^0_3\! \left( 1_Q, 3_g, 4_g \right) 
\left|\cM^{0}_3  \! \left( \widetilde{ (13) }_Q, \widetilde{ (34) }_g, 2_{\bar{Q}} \right) \right|^2
 {\cF}_2^{(3)} \!  \left( \widetilde{ k_{13} }, \widetilde{ k_{34} }, k_2 \right) 
\nonumber \\[0.5ex]
& & \qquad \quad {} + d^0_3\! \left( 1_Q, 4_g, 3_g \right) 
\left|\cM^{0}_3  \! \left( \widetilde{ (14) }_Q, \widetilde{ (43) }_g, 2_{\bar{Q}} \right) \right|^2
{\cF}_2^{(3)}  \!  \left( \widetilde{ k_{14} },\widetilde{ k_{43} }, k_2 \right) 
\nonumber \\[0.5ex]
& & \qquad \quad {} + d^0_3\! \left(2_{\bar{Q}}, 3_g, 4_g \right)
\left|\cM^{0}_3  \! \left( 1_Q , \widetilde{ (43) }_g, \widetilde{(32)}_{\bar{Q}} \right) \right|^2
{\cF}_2^{(3)}  \! \left(  k_1, \widetilde{ k_{43}}, \widetilde{ k_{32} } \right) 
 \nonumber \\
& & \qquad \quad {} + d^0_3\! \left(2_{\bar{Q}}, 4_g, 3_g \right)
\left|\cM^{0}_3  \! \left( 1_Q , \widetilde{ (34) }_g, \widetilde{(42)}_{\bar{Q}} \right) \right|^2
{\cF}_2^{(3)}  \! \left( k_1, \widetilde{ k_{34}}, \widetilde{ k_{42} } \right)
\bigg] \nonumber \\[1ex]
& & \qquad {} - \frac{1}{N_c} \bigg[ A^0_3 \! \left( 1_Q , 3_g , 2_{\bar{Q}} 
\right) 
\left|\cM^{0}_3  \! \left( \widetilde{ (13) }_Q, 4_g, \widetilde{(32)}_{\bar{Q}} \right) \right|^2
{\cF}_2^{(3)}  \! \left( \widetilde{ k_{13}}, k_4, \widetilde{ k_{32} } \right)
\nonumber \\
& & \qquad \quad {} + A^0_3 \! \left( 1_Q , 4_g , 2_{\bar{Q}} \right) 
\left|\cM^{0}_3  \! \left( \widetilde{ (14) }_Q, 3_g, \widetilde{(42)}_{\bar{Q}} \right) \right|^2
{\cF}_2^{(3)}  \! \left( \widetilde{ k_{14}}, k_3, \widetilde{ k_{42} } \right)
\bigg] \bigg\},      
%
\end{eqnarray}

\begin{eqnarray}\label{sub:ggb2}
\lefteqn{ d \sigma^{S,b,2, Q \bar{Q} g g}_\mss{NNLO} =
 \einhalb \, {\cN} 
\, d \Phi_4 ( k_1, k_2, k_3, k_4; q )}
 \quad \nonumber \\[0.5ex]
& & {} \times \bigg\{ \, N_c \bigg[
A^0_4 \! \left ( k_1, k_3, k_4, k_2 \right) \left|{\cM}^0_2\left(\widetilde{(134)}_Q, \widetilde{(342)}_{\bar Q}\right)\right|^2
{\cF}_2^{(2)} \! \left( \widetilde{ k_{134} }, \widetilde{ k_{342} } \right)  \nonumber \\[0.5ex]
 & & {}  +  A^0_4 \! \left ( k_1, k_4, k_3, k_2 \right) \left|{\cM}^0_2\left(\widetilde{(143)}_Q, \widetilde{(432)}_{\bar Q}\right)\right|^2
{\cF}_2^{(2)} \! \left( \widetilde{ k_{143} }, \widetilde{ k_{432} } \right) \bigg]\nonumber \\[0.5ex]
& & {} - \frac{1}{ N_c } \tilde{ A }^0_4 \! \left( 1_Q, 3_g, 4_g, 2_{\bar{Q}} \right) 
\left|{\cM}^0_2\left(\widetilde{(134)}_Q, \widetilde{(342)}_{\bar Q}\right)\right|^2
{\cF}_2^{(2)} \! \left( \widetilde{ k_{134} }, \widetilde{ k_{342} } \right) \bigg\} \, ,
\end{eqnarray}

\begin{eqnarray}\label{sub:ggb1}
\lefteqn{ d \sigma^{S,b,1, Q \bar{Q} g g}_\mss{NNLO} =
- \einhalb\,{\cN} 
\, d \Phi_4 ( k_1, k_2, k_3, k_4; q )}
 \quad
\nonumber \\[0.5ex]
& & {} \times \bigg\{ \, N_c \bigg[
 d^0_3\! \left( 1_Q, 3_g, 4_g \right) 
A^{0}_3  \! \left( \widetilde{ (13) }_Q, \widetilde{ (34) }_g, 2_{\bar{Q}} \right) 
\left|\cM_2^0\left(\widetilde{(\widetilde{(13)}\widetilde{(34)})_Q},\widetilde{(\widetilde{(34)}2)_{\bar Q}}\right)\right|^2
\cF_2^{(2)}\left(\widetilde{k_{\widetilde{(13)}\widetilde{(34)}}}, \widetilde{k_{\widetilde{(34)}2}}\right)
\nonumber \\[0.5ex]
& & {} + d^0_3\! \left( 1_Q, 4_g, 3_g \right) 
A^{0}_3  \! \left( \widetilde{ (14) }_Q, \widetilde{ (43) }_g, 2_{\bar{Q}} \right) 
\left|\cM_2^0\left(\widetilde{(\widetilde{(14)}\widetilde{(43)})_Q},\widetilde{(\widetilde{(43)}2)_{\bar Q}}\right)\right|^2
\cF_2^{(2)}\left(\widetilde{k_{\widetilde{(14)}\widetilde{(43)}}}, \widetilde{k_{\widetilde{(43)}2}}\right)
\nonumber \\[0.5ex]
& &  {} + d^0_3\! \left(2_{\bar{Q}}, 3_g, 4_g \right) A^{0}_3 \! \left( 
1_Q , \widetilde{ (43) }_g, \widetilde{(32)}_{\bar{Q}} \right)
\left|\cM_2^0\left(\widetilde{(1\widetilde{(43)})_Q}, \widetilde{(\widetilde{(43)}\widetilde{(32)})_{\bar Q}}\right)\right|^2
\cF_2^{(2)}\left(\widetilde{k_{1\widetilde{(43)}}}, \widetilde{k_{\widetilde{(43)}\widetilde{(32)}}}\right) 
\nonumber \\[0.5ex]
& &  {} + d^0_3\! \left(2_{\bar{Q}}, 4_g, 3_g \right) A^{0}_3 \! \left( 
1_Q , \widetilde{ (34) }_g, \widetilde{(42)}_{\bar{Q}} \right)
\left|\cM_2^0\left(\widetilde{(1\widetilde{(34)})_Q}, \widetilde{(\widetilde{(34)}\widetilde{(42)})_{\bar Q}}\right)\right|^2
\cF_2^{(2)}\left(\widetilde{k_{1\widetilde{(34)}}}, \widetilde{k_{\widetilde{(34)}\widetilde{(42)}}}\right)\bigg]
\nonumber \\[0.5ex]
& & {} - \frac{1}{ N_c } \bigg[  A^0_3 \! \left( 1_Q , 3_g , 2_{\bar{Q}} \right) A^0_3 
\! 
\left( \widetilde{(13)}_Q , 4_g , \widetilde{(32)}_{\bar{Q}} \right)
\left|{\cM}^0_2\left(\widetilde{(\widetilde{(13)}4)}_Q, \widetilde{(4\widetilde{(32)})}_{\bar{Q}}\right)\right|^2
{\cF}_2^{(2)} \! \left(\widetilde{k_{\widetilde{(13)}4}}, \widetilde{k_{4\widetilde{(32)}}}\right)
\nonumber \\[0.5ex]
& & {} + A^0_3 \! \left( 1_Q , 4_g , 2_{\bar{Q}} \right) 
A^0_3 \! \left( \widetilde{ (14) }_Q , 3_g , \widetilde{ (42) }_{\bar{Q}} \right)
\left|{\cM}^0_2\left(\widetilde{(\widetilde{(14)}3)}_Q, \widetilde{(3\widetilde{(42)})}_{\bar{Q}} \right)\right|^2
{\cF}_2^{(2)} \! \left(\widetilde{k_{\widetilde{(14)}3}}, \widetilde{k_{3\widetilde{(42)}}}\right) \bigg]\bigg\} \, . \nonumber \\[0.5ex]
\end{eqnarray}
The tree-level massive quark gluon antenna function $d_3^0$ is given in \cite{GehrmannDeRidder:2009fz}.
The four-parton $Q{\bar Q}gg$ antenna functions $A_4^0$ and ${\tilde A}_4^0$, which were derived in \cite{Bernreuther:2013uma},
govern the color-ordered and non-ordered (photon-like) emission between a pair of massive radiator quarks, respectively.
The mapped momenta denoted by a tilde and double tilde in \eqref{sub:gga} -- \eqref{sub:ggb1} are obtained from $3\to 2$, $4\to 2$,
and two iterated $3\to 2$ mappings, respectively, in completely analogous fashion as in the $Q{\bar Q}q{\bar q}$ case; cf. appendix~\ref{sec:AppA}.

 The remarks on the angular correlations due to gluon splitting made in the second paragraph below eq.~\eqref{eq:Sb1qqbar} apply also here, where the 
 angular correlations are due to $g\to g g$. These correlations are present only in the leading-color part of 
  the unsubtracted differential cross section \eqref{eq:unsigQQgg} and in the  leading-color part of the subtraction term \eqref{sub:ggb2}.
  In analogy to the $Q{\bar Q}q{\bar q}$ case we evaluate these leading-color terms  
  for each set of momenta $k_1,k_2,k_3,k_4$
      also for  $k_1, k_2, k_{3r}, k_{4r}$ and take the average. 
      We sampled the phase space   for this final state
      in regions where $k_3\cdot k_4/s \lesssim 10^{-8}$ and checked that  
      the resulting subtraction term is a very good approximation to the squared matrix element in all single and double unresolved limits.

\subsection{Real-virtual corrections}
\label{suse:revirc}
In this subsection we outline  how to compute the order $\alpha_s^2$  contribution of the $Q{\bar Q}g$ final state to 
 the differential massive quark-pair production cross section  with the antenna subtraction method;
that is, the second term  on the right-hand side of the
  first line of \eqref{eq:sub2NNLO}. 
\subsubsection*{Unsubtracted real-virtual cross section:}
  This contribution involves the interference of the tree-level and one-loop 
 $Q{\bar Q}g$ final-state amplitude. Using the conventions of \eqref{eq::M03_S2QQgA}
  the unsubtracted $\cO(\alpha_s^2)$ correction to the cross section, summed over colors and summed/averaged over spins,
   is given by 
\begin{eqnarray}
d \sigma^{RV,Q \bar{Q} g}_\mss{NNLO} & = & 
     {\cN} C(\epsilon) 
d \Phi_3 ( k_1, k_2, k_3 ; q ) \,
\cF^{(3)}_2 \! \left(  k_1, k_2, k_3 \right)
	\nonumber \\[0.5ex]
	& & {} \times \left[
	N_c \,
	\delta
	\cM^{1,{\rm lc}}_3
	\! \left( 1_Q, 3_g, 2_{\bar{Q}} \right) 
	- \frac{1}{N_c} \, 
	\delta
	\cM^{1,{\rm sc}}_3
	\! \left( 1_Q, 3_g, 2_{\bar{Q}} \right) 
	\right.
	+ n_f \delta
	\cM^{1,f}_3
	\! \left( 1_Q, 3_g, 2_{\bar{Q}} \right) \nonumber \\
	& & {}  +
	\delta
	\cM^{1,F}_3
	\! \left( 1_Q, 3_g, 2_{\bar{Q}} \right) 
	+
	\delta
	\cM^{1,{\rm tr}}_3
	\! \left( 1_Q, 3_g, 2_{\bar{Q}} \right) 	
	\bigg] \, .
\label{xs:unQQg}
\end{eqnarray}
The factors   $C(\epsilon)$  and $\cN$ are defined in \eqref{eq:defCeps} and \eqref{eq:defN}, respectively,
 and we have introduced the shorthand notation 
\begin{equation}\label{eq:defdM13}
\delta \cM^{1,X}_3  
\! \left( 1_Q, 3_g, 2_{\bar{Q}} \right) 
= 2\,\mbox{Re} \left[ \left( 
\cM^0_3  \! \left( 1_Q, 3_g, 2_{\bar{Q}} \right)  \right)^\ast
\cM^{1,X}_3 \! \left( 1_Q, 3_g, 2_{\bar{Q}} \right) \right],
\end{equation}
with $X \in \{ {\rm lc}, {\rm sc}, f, F, {\rm tr}\}$. We recall that $\cM^{1}_3$ is the renormalized one-loop amplitude.  
The analytic computation of 
\eqref{xs:unQQg}, which was first performed in  
\cite{Bernreuther:1997jn,Brandenburg:1997pu,Rodrigo:1997gy,Rodrigo:1999qg,Nason:1997tz,Nason:1997nw},
 is standard by now. We recall from section~\ref{susec:nlo} that the triangle term $\delta \cM^{1,{\rm tr}}_3$, which was analyzed first in 
 \cite{Hagiwara:1990dx}, is an IR finite and non-universal QCD correction.
 The other contributions to the unsubtracted cross section contain explicit IR poles (single and double poles in $1/\epsilon$).
  In addition, the phase-space integration of \eqref{xs:unQQg} in the region 
  where the external gluon becomes soft, leads to additional IR singularities.
  Both types of singularities must be subtracted with appropriate terms 
  in order that the integration over the three-parton phase space can be performed 
   numerically in four dimensions. 

%
\subsubsection*{Subtraction of explicit infrared poles:}
 The explicit IR poles in
\eqref{xs:unQQg} are removed by adding the 
subtraction terms \eqref{eq:SQQqqa} and \eqref{sub:gga}, integrated over the phase-space of one unresolved parton:
\begin{eqnarray}
d \sigma^{T, a, Q \bar{Q} g}_\mss{NNLO} 
 & = & 
- \int_1 d \sigma^{S,a, Q \bar{Q} g g}_\mss{NNLO}
- \sum_q \int_1 d \sigma^{S,a, Q \bar{Q} q \bar{q} }_\mss{NNLO} 
\nonumber \\ 
& = & -
\cN C( \epsilon ) d\Phi_3 ( k_1, k_2, k_3 ; q ) \,
\cF^{(3)}_2 \! \left(  k_1, k_2, k_3 \right)
\nonumber \\[0.5ex]
& & {} \times \bigg[ \, \frac{N_c}{2} \Big( \, 
  \cD^0_3\! \left( \epsilon, \mu^2 / k^2_{13} ; z_{13} \right) 
+ \cD^0_3\! \left( \epsilon, \mu^2 / k^2_{23} ; z_{23} \right) \Big) -
\frac{1}{N_c} \cA^0_3 
\! \left( \epsilon, \mu^2 / k^2_{12} ; y_{12} \right) 
\nonumber \\
& & {} + \frac{n_f}{2} \, \Big( \,
  \cE^0_3\! \left( \epsilon , \mu^2 / k_{13}^2 ; z_{13}  \right) 
+ \cE^0_3\! \left( \epsilon , \mu^2 / k_{23}^2 ; z_{23}  \right)
\Big) \bigg]
\big|
\cM^0_3  \! \left( 1_Q, 3_g, 2_{\bar{Q}} \right) 
\!\big|^2 .
\label{sub::rv::a::int}
\end{eqnarray}
The antenna functions $\cD^0_3$ and $\cE^0_3$, which are the integrated versions of the tree-level
 antenna functions $d_3^0$ and $E_3^0$, respectively, are given in  \cite{GehrmannDeRidder:2009fz,Abelof:2011jv}.
 The  poles in $\epsilon$ of these functions and of $\cA^0_3$  cancel the explicit IR poles  in \eqref{xs:unQQg}.
 The kinematic invariants that appear in the arguments of these functions are 
 \begin{equation}\label{eq:kininvRV}
  k_{ij}^2=(k_i+k_j)^2, \quad z_{ij}=\frac{m_Q}{\sqrt{k_{ij}^2}}, \quad \beta_{ij}=\sqrt{1-4 z_{ij}^2}, \quad 
  y_{ij}=\frac{1-\beta_{ij}}{1+\beta_{ij}} \, .
 \end{equation}

\subsubsection*{One-loop single-unresolved subtraction term:}
The singular behavior of \eqref{xs:unQQg} is mimicked in the limit where the external gluon becomes
unresolved  by the following subtraction term:
\begin{eqnarray}
 d \sigma^{T,b,Q \bar{Q} g}_\mss{NNLO} & = & 
\cN C( \epsilon )  d\Phi_3 ( k_1, k_2, k_3 ; q ) \,
\cF^{(2)}_2 \!  \left( \widetilde{k_{13}} , \widetilde{k_{32}}  \right) 
\nonumber \\
& & {} \times \bigg(
N_c \bigg[ 
  A^1_3 \! \left( 1_Q, 3_g, 2_{\bar{Q}} \right) 
\Big|
\cM^0_2  \! 
\left(  \widetilde{ ( 13 ) }_Q, \widetilde{ ( 32 ) }_{\bar{Q}} \right) 
\!\Big|^2 
\nonumber \\
& & \hspace{10ex} \quad { }
+ A^0_3 \! \left( 1_Q, 3_g, 2_{\bar{Q}} \right) 
\delta \cM^1_2  \! 
\left(  \widetilde{ ( 13 ) }_Q, \widetilde{ ( 32 ) }_{\bar{Q}} \right) 
\bigg]
\nonumber \\
& & \quad { } - \frac{1}{N_c} \bigg[ 
\tilde{A}^1_3 \! \left( 1_Q, 3_g, 2_{\bar{Q}} \right) 
\Big|
\cM^0_2  \! 
\left(  \widetilde{ ( 13 ) }_Q, \widetilde{ ( 32 ) }_{\bar{Q}} \right) 
\!\Big|^2 
\nonumber \\
& & \hspace{10ex} \quad { }
+
A^0_3 \! \left( 1_Q, 3_g, 2_{\bar{Q}} \right) 
\delta\cM^{1}_2 \! 
\left(  \widetilde{ ( 13 ) }_Q, \widetilde{ ( 32 ) }_{\bar{Q}} \right) 
\bigg] 
\nonumber \\
& & \; \;
+ \left( n_f \, \hat{A}^1_{3,f} \! \left( 1_Q, 3_g, 2_{\bar{Q}} \right)
+ \hat{A}^1_{3,F} \! \left( 1_Q, 3_g, 2_{\bar{Q}} \right) \right)
\Big|
\cM^0_2  \! 
\left(  \widetilde{ ( 13 ) }_Q, \widetilde{ ( 32) }_{\bar{Q}} \right) 
\!\Big|^2 
\bigg). \quad 
\label{QQg::sub::b}
\end{eqnarray}
The massive one-loop antenna functions $A^1_{3}$, $\tilde{A}^1_{3}$,
$\hat{A}^1_{3,f}$, and $\hat{A}^1_{3,F}$ were determined in \cite{Dekkers:2014hna}.
 The unintegrated  tree-level massive quark-antiquark antenna $A^0_3$
  was already introduced in section~\ref{susec:nlo}.
 The Born times one-loop interference term $\delta\cM^{1}_2$ is defined in \eqref{eq:delLONLO2}.
%
%
\subsubsection*{Compensation term for oversubtracted poles:}
In certain regions of phase space, the subtraction terms \eqref{sub::rv::a::int}
and \eqref{QQg::sub::b} exhibit IR singularities that do not 
coincide with respective singularities in the unsubtracted real-virtual cross section
\eqref{xs:unQQg}. In order to remove these spurious singularities one has to
introduce an additional subtraction term that is given by \cite{Dekkers:2014hna}
\begin{equation} \label{eq:TcQQg}
  d \sigma^{T,c,Q \bar{Q} g}_\mss{NNLO} = 
- \int_1 d \sigma^{S,b,1,Q\bar{Q} g g }_\mss{NNLO} 
- \int_1 d \sigma^{S,b,1,Q\bar{Q} q \bar{q}}_\mss{NNLO}\, \, .
\end{equation}
 The integrands are given in \eqref{eq:Sb1qqbar}
 and \eqref{sub:ggb1}, respectively. The integration over the phase space of one unresolved parton yields
\begin{eqnarray}
 d \sigma^{T,c,Q \bar{Q} g}_\mss{NNLO} 
& = & 
\cN C( \epsilon )  d\Phi_3 ( k_1, k_2, k_3 ; q ) \,
\left| 
\cM^0_2 \!
\left( \widetilde{ ( 13 ) }_Q , \widetilde{ ( 32 ) }_{\bar{Q}} \right) 
\right|^2 
\cF^{(2)}_2 \!  \left( \widetilde{p_{{13}}} , \widetilde{p_{32}}  \right) 
\nonumber \\[0.5ex]
& & {} \times \bigg[ \, \frac{N_c}{2} \Big( \, 
  \cD^0_3\! \left( \epsilon, \mu^2 / k^2_{13} ; z_{13} \right) 
+ \cD^0_3\! \left( \epsilon, \mu^2 / k^2_{23} ; z_{23} \right) \Big) -
\frac{1}{N_c} \cA^0_3 
\! \left( \epsilon, \mu^2 / k^2_{12} ; y_{12} \right) 
\nonumber \\
& & {} + \frac{n_f}{2} \, \Big( \,
  \cE^0_3\! \left( \epsilon , \mu^2 / k_{13}^2 ; z_{13}  \right) 
+ \cE^0_3\! \left( \epsilon , \mu^2 / k_{23}^2 ; z_{23}  \right)
\Big) \bigg]
A^0_3 \! \left( 1_Q, 3_g, 2_{\bar{Q}} \right)  \, .
\label{QQg::sub::c::1}
\end{eqnarray}
This counterterm serves a twofold purpose. In the region of the three-parton phase space where
the gluon is resolved, subtracting \eqref{sub::rv::a::int}
from the unsubtracted real-virtual cross section \eqref{xs:unQQg} leads to a finite integrand. In these
regions, $d\sigma^{T,c,Q \bar{Q} g}_\mss{NNLO}$ cancels the explicit IR poles
of $d\sigma^{T,b,Q \bar{Q} g}_\mss{NNLO}$ contained in the one-loop antenna
functions and $\delta\cM^1_2$. 
On the other hand, in the region where the gluon is
unresolved, $d\sigma^{T,b,Q \bar{Q} g}_\mss{NNLO}$ correctly approximates all
(explicit and implicit) infrared singularities of 
\eqref{xs:unQQg}. In this region, the sum of $d\sigma^{T,a,Q \bar{Q}
g}_\mss{NNLO}$ and $d\sigma^{T,c,Q \bar{Q} g}_\mss{NNLO}$ yields a finite 
contribution.
%
\subsubsection*{Summary:}
 Combining eqs.~\eqref{xs:unQQg}, \eqref{sub::rv::a::int},
\eqref{QQg::sub::b}, and \eqref{QQg::sub::c::1}
yields an expression that is free of (explicit and implicit) singularities
in the entire three-parton phase space in $D = 4$ dimensions:
\begin{equation}
 \int_{\Phi_3} \left[ 
  d \sigma^{RV,Q \bar{Q} g}_\mss{NNLO}
- d \sigma^{T, a, Q \bar{Q} g }_\mss{NNLO} 
- d \sigma^{T, b, Q \bar{Q} g }_\mss{NNLO} 
- d \sigma^{T, c, Q \bar{Q} g }_\mss{NNLO} 
\right]_{ \epsilon = 0 }
= \mbox{finite}.
\end{equation}
We recall that the terms $d\sigma^{T, a, Q \bar{Q} g
}_\mss{NNLO}$ and $d\sigma^{T, c, Q \bar{Q} g }_\mss{NNLO}$ are counterbalanced
by the double-real subtraction terms $d\sigma^{S, a, Q \bar{Q} ij}_\mss{NNLO}$
and $d\sigma^{S, b, 1, Q \bar{Q} ij}_\mss{NNLO}$ $(ij=gg, q{\bar q})$, respectively, that were
 defined in section~\ref{suse:RR}. Hence, only the integrated form of $d\sigma^{T, b, Q\bar{Q} g }_\mss{NNLO}$ 
  has to be added back to the double virtual
contribution that will be discussed in the next subsection.
%
\subsection{Double virtual corrections}
\label{suse:doubVV}
Finally, we discuss  how to compute the order $\alpha_s^2$  contribution of the $Q{\bar Q}$ final state to 
 the differential massive quark-pair production cross section  within the antenna framework,
that is, the sum of the three terms in the 
  second line of \eqref{eq:sub2NNLO}. 
\subsubsection*{Unsubtracted real-virtual cross section:}

The renormalized one-loop and 2-loop $Q{\bar Q}$ matrix elements 
 defined in \eqref{eq::M02_S2QQ}
 yield the following
$\cO( \alpha_s^2 )$ correction to the differential cross section:
\begin{eqnarray}
 d \sigma^{VV,Q \bar{Q} }_\mss{NNLO} & = & 
\frac{\left( \bar{C} \! \left( \epsilon \right) \right)^2}{8 s}
\left( \frac{ \alpha_s }{ 2 \pi } \right)^2
 N_c \,
 d \Phi_2\!\left( k_1, k_2 ; q \right)
 \cF^{(2)}_2 \!  \left( k_1 , k_2 \right)
    \Big\{ 4 C^2_F \,
      \big| \cM^1_2 \! \left( 1_Q, 2_{\bar{Q}} \right) \! \big|^2
    \nonumber \\
    & & {} 
    + 2 {\rm Re}[\cM^{0\ast}_2 \! \left( 1_Q, 2_{\bar{Q}} \right) 
    \cM^2_2 \! \left( 1_Q, 2_{\bar{Q}} \right) ]
   \Big\}.
\label{VV:xs}
\end{eqnarray}
Summation over all colors and spins is understood. The spin averaging factor $1/4$ for unpolarized $e^-e^+$
 collisions is included in \eqref{VV:xs}. The factor $\bar{C}(\epsilon)$ is defined in  \eqref{eq:defCeps}.
 The real and imaginary parts of the one-loop vertex functions $VQ{\bar Q}$ $(V=\gamma, Z)$, up to and including terms of order $\epsilon$,
  and the corresponding two-loop vertex functions  to order  $\epsilon^0$ were computed in 
  \cite{Bernreuther:2004ih,Bernreuther:2004th,Bernreuther:2005rw}; cf. also \cite{Gluza:2009yy}.
  With these vertex functions, \eqref{VV:xs} can be computed in straightforward fashion. The last
  term on the right-hand side of \eqref{VV:xs} can be decomposed into different color structures, that is, leading and subleading color contributions, terms that involve a massless and massive quark 
   loop in the gluon vacuum polarization tensor, and triangle contributions summed over all quark flavors. These triangle contributions, which are finite \cite{Bernreuther:2005rw}, are non-universal QCD
   corrections to the leading-order cross section.

\subsubsection*{Subtraction term:}

Recalling the subtraction terms that were introduced above, those that remain to be 
counterbalanced are $d\sigma^{T, b, Q\bar{Q} g }_\mss{NNLO}$ (cf. eq.~\eqref{QQg::sub::b}) and 
 $d\sigma^{S,b,2, Q \bar{Q} ij}$ $(ij=q{\bar q}, gg)$ (cf. eq.~\eqref{eq:Sb2qqbar} and~\eqref{sub:ggb2}).
 They have to be integrated over the unresolved one-parton, respectively two-parton phase space
  in order to serve as counterterm for subtracting the IR poles in 
   $\epsilon$ of the double-virtual correction \eqref{VV:xs}.
  We get 
\begin{eqnarray}
\lefteqn{
\int_1 d \sigma^{T,b, Q \bar{Q} g }_\mss{NNLO} 
+ \int_2 d \sigma^{S,b,2, Q \bar{Q} g g}_\mss{NNLO}
+ \sum_q \int_2 d \sigma^{S,b,2, Q \bar{Q} q \bar{q} }_\mss{NNLO} 
} \quad 
\nonumber \\
& = & 
\frac{\left( \bar{C} \! \left( \epsilon \right) \right)^2}{8 s}
\left( \frac{ \alpha_s }{ 2 \pi } \right)^2
\left( N_c^2 - 1 \right)
 d \Phi_2\!\left( k_1, k_2 ; q \right)
 \cF^{(2)}_2 \!  \left( k_1 , k_2 \right)
\nonumber \\
& & {} \times \bigg\{ N_c \bigg[ \Big(
\cA^0_4 \! \left( \epsilon , \mu^2 /s ; y \right) 
+ \cA^1_3 \! \left( \epsilon, \mu^2 /s ; y \right) 
\Big)
\big|
\cM^0_2  \! 
\left(  1_Q, 2_{\bar{Q}} \right) 
\!\big|^2 
\nonumber \\
& & \hspace{10ex} {} + 
\cA^0_3 \! \left( \epsilon, \mu^2 /s ; y \right) 
 \delta\cM^1_2  \! 
\left(  1_Q, 2_{\bar{Q}} \right) 
\bigg]
\nonumber \\
& & {} - \frac{1}{ N_c } \left[ \left(
\frac{1}{2} \, 
  \tilde{\cA}^0_4 \! \left( \epsilon , \mu^2 /s ; y \right)  
+ \tilde{\cA}^1_3 \! \left( \epsilon, \mu^2 /s ; y \right) 
\right)
\big|
\cM^0_2  \! 
\left(  1_Q, 2_{\bar{Q}} \right) 
\!\big|^2 
\right.
\nonumber \\
& & \hspace{10ex} {} +
\cA^0_3 \! \left( \epsilon, \mu^2 /s ; y \right) 
 \delta \cM^1_2  \! 
\left(  1_Q, 2_{\bar{Q}} \right) 
\bigg]
\nonumber \\
& & {} + 2 \, T_R \, n_f \left(
 \cB^0_4 \! \left( \epsilon, \mu^2 /s ; y \right) 
+ \hat{\cA}^1_{3,f} \! \left( \epsilon, \mu^2 /s ; y \right) 
\right)
\big|
\cM^0_2 \! 
\left(  1_Q, 2_{\bar{Q}} \right) 
\!\big|^2 
\nonumber \\
& & {} + 2 \, T_R \,
\hat{\cA}^1_{3,F} \! \left( \epsilon, \mu^2 /s ; y \right) 
\big|
\cM^0_2 \! 
\left(  1_Q, 2_{\bar{Q}} \right) 
\!\big|^2 
\bigg\} \, .
\label{VV::sub}
\end{eqnarray}
 The variable $y$ is defined in \eqref{eq:defybet}. The integrated antenna functions 
 $\cB^0_4$, $\cA^0_4$, and $\tilde{\cA}^0_4$ were computed in 
\cite{Bernreuther:2011jt,Bernreuther:2013uma} and
$\cA^1_3$, $\tilde{\cA}^1_3$, $\hat{\cA}^1_{3,f}$, and   $\hat{\cA}^1_{3,F}$ were
 determined in \cite{Dekkers:2014hna}.

The subtraction term \eqref{VV::sub} has to be added to \eqref{VV:xs}.
 In the sum all IR poles cancel in analytic fashion.
\begin{equation}
 \int_{\Phi_2} \left[ 
  d \sigma^{VV,Q \bar{Q} }_\mss{NNLO}
+ \int_1 d \sigma^{T,b, Q \bar{Q} g }_\mss{NNLO} 
+ \int_2 d \sigma^{S,b,2, Q \bar{Q} g g}_\mss{NNLO}
+ \sum_q  \int_2 d \sigma^{S,b,2, Q \bar{Q} q \bar{q} }_\mss{NNLO} 
\right]_{ \epsilon = 0 }
= \mbox{finite}.
\end{equation}
 After summing these terms and after
analytic cancellation of the IR poles, one can  take the limit $\epsilon 
\to 0$ and perform the remaining integration over the two-parton phase space in four dimensions.
  
  \section{Results for top-quark pair production}
  \label{sec:restt}
 In this section we present our  numerical results for the total $\ttbar$ cross section  and for several distributions, 
  including the top-quark forward-backward asymmetry above the $\ttbar$ threshold at order $\as^2$. We use the input values
  $m_W=80.385$ GeV,  $m_Z=91.1876$ GeV, and $\Gamma_Z=2.4952$ GeV \cite{Agashe:2014kda}. We use  $m_t=173.34$ GeV for the top-quark mass in the on-shell scheme. 
   The other quarks are taken to be massless. 
  The sine of the weak mixing angle,  $s_W$,  is fixed by $s^2_W=1-m_W^2/m_Z^2$. For computing the electroweak couplings we use the so-called $G_\mu$ scheme
 (cf., for instance, \cite{Beenakker:1991ca})  where the electromagnetic coupling is given by $\alpha=\sqrt{2}G_\mu m_W^2 s_W^2/\pi= 7.5624 \times 10^{-3}$ with $G_\mu=1.166379\times 10^{-5}$ ${\rm GeV}^{-2}$.
  The  running of the ${\overline{\rm MS}}$ QCD coupling $\alpha_s(\mu)$ is determined in $f=6$ flavor QCD from the input value $\alpha_s^{(f=5)}(\mu=m_Z)=0.118$. 
   In this section $\mu$ refers to the  renormalization scale.
  
  Because we work to lowest order in the electroweak couplings each of the various contributions $d\sigma^{(i,j)}$ to the differential $Q{\bar Q}$ cross section at order $\alpha_s^2$
   discussed in section~\ref{sec:lonlo} and~\ref{sec:nnlo} is given by the sum of an s-channel $\gamma$ and $Z$-boson contribution  and a $\gamma Z$ interference term. 
   The $d\sigma^{(i,j)}$ have the structure
   \begin{equation}\label{eq:strucsig} 
   d\sigma^{(i,j)} = \sum\limits_{a=\gamma, Z, \gamma Z} K_a^{(j)} \: L^{\mu\nu}_a H^{(i,j)}_{a,\mu\nu} \: d\Phi_j\, .
   \end{equation}
Here $L^{\mu\nu}_a$ denote the lepton tensors (with the boson propagators included) and $H^{(i,j)}_{a,\mu\nu}$ the antenna-subtracted, i.e., IR finite parton tensors
of order $\alpha_s^j$. Thus the Lorentz contractions and the phase-space integration in \eqref{eq:strucsig} can be done in $D=4$ dimensions.
The first index $i$ in the superscript $(i,j)$ labels the final state, i.e., $i= Q{\bar Q}$, $Q{\bar Q}g$, $Q{\bar Q}gg$, $Q{\bar Q}q {\bar q}$. 
 The factors  $K_a^{(j)}$ contain the 
 electroweak couplings, the flux factor, and the $e^-e^+$ spin-averaging factor. (The label $j$  in $K_a^{(j)}$ is 
  necessary because some second-order matrix elements 
   involve also terms that contain the electroweak couplings of $q\neq Q$, see below.) In this work we consider unpolarized  $e^-e^+$ collisions.
  We separate each contribution $(i,j)$ on the right-hand side of  \eqref{eq:strucsig} into a  parity-even and -odd term.
  
  The two-loop $t{\bar t}$ matrix elements and the integrated antenna subtraction terms discussed in section~\ref{sec:nnlo} contain 
  harmonic polylogarithms (HPL) \cite{Remiddi:1999ew}. We evaluate them with the codes of refs.~\cite{Gehrmann:2001pz,Maitre:2005uu}.
  The integrated antenna functions $\cA^1_3$, $\tilde{\cA}^1_3$ that appear in \eqref{VV::sub} are expressed in terms of HPL and  cyclotomic harmonic polylogarithms 
   \cite{Blumlein:2009ta,Ablinger:2011te,Ablinger:2013cf}. We evaluate them numerically by using the integral representation of these functions.
  
  For center-of-mass (c.m.) energies $\sqrt{s} > 4 m_t$, four-top production, i.e., $t{\bar t} t{\bar t}$ production occurs. The order $\alpha_s^2$  cross section 
  of this process is  infrared-finite. It makes only a small contribution to the inclusive $\ttbar$ cross section. Moreover, the $t{\bar t} t{\bar t}$ final state 
   has a distinct signature  and could be experimentally distinguished from $\ttbar$ final states. Below we consider c.m. energies  $\sqrt{s} \lesssim 4 m_t$.

\subsection{Cross section and distributions}
\label{suse:xsecdist}

\begin{figure}[tbh!]
 \centering
 \includegraphics[width=14cm,height=10cm]{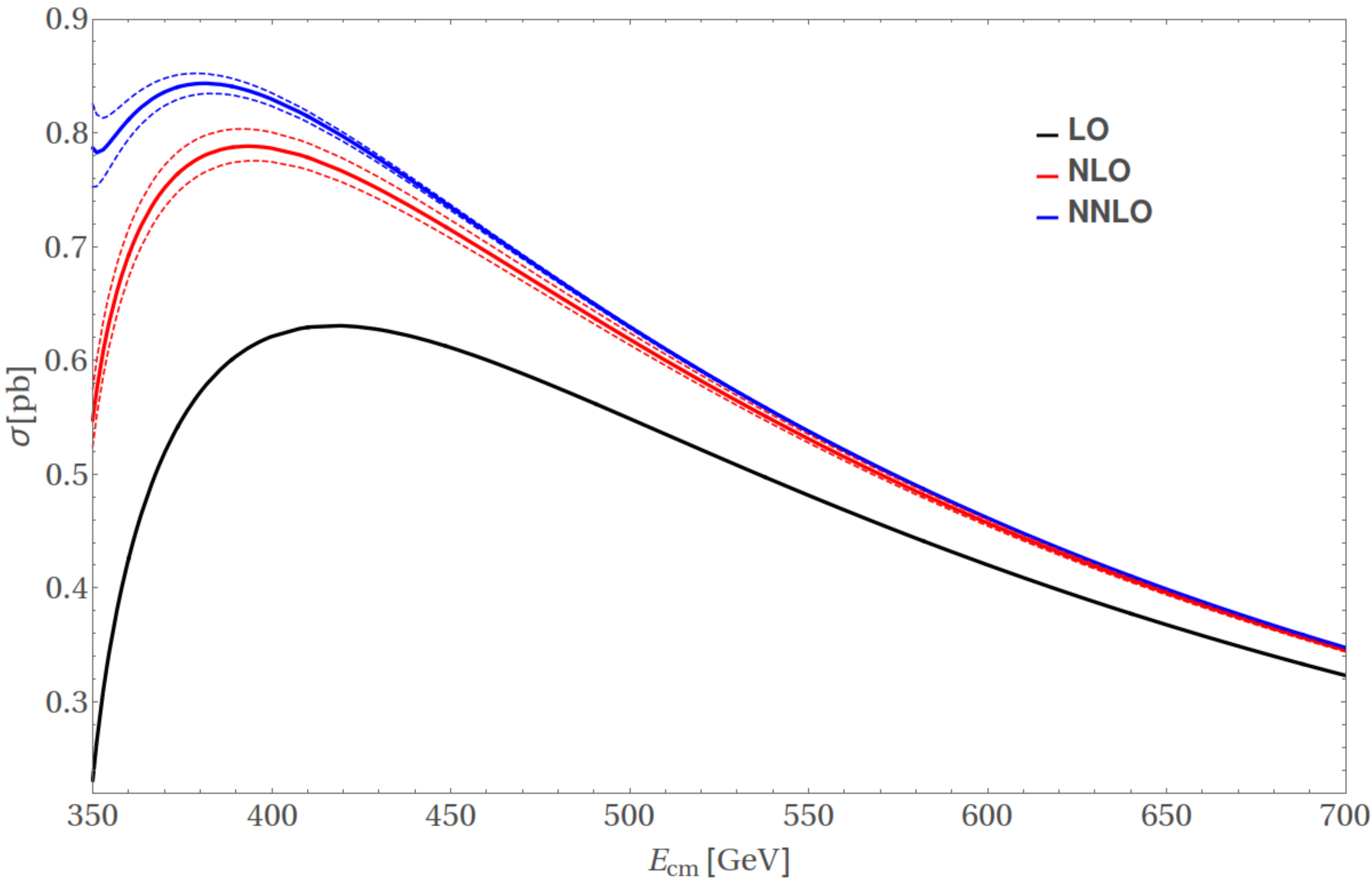}
 \caption{The total $\ttbar$ cross section at LO, NLO, and NNLO QCD as a function of the c.m. energy.
  The solid lines that refer to  $\sigma_{\rm NLO}$ and $\sigma_{\rm NNLO}$
  correspond to the choice $\mu=\sqrt{s}$, the dashed lines correspond to 
  $\mu=\sqrt{s}/2$ and $2\sqrt{s}$.  }
 \label{fig:sigtt}
 \end{figure}

 Figure~\ref{fig:sigtt} shows our result for  the $\ttbar$ cross section at LO, NLO, and NNLO QCD for c.m. energies below the four-top threshold.
  In the case of $\sigma_{\rm NLO}$ and $\sigma_{\rm NNLO}$ the solid lines correspond to the choice $\mu=\sqrt{s}$ for the renormalization scale.
  Uncertainties due to undetermined higher-order corrections are conventionally estimated by varying $\mu$ between $\sqrt{s}/2$ and $2\sqrt{s}$.
  The upper and lower dashed lines correspond to these scale variations.
 
\begin{figure}[tbh!]
 \centering
 \includegraphics[width=14cm,height=10cm]{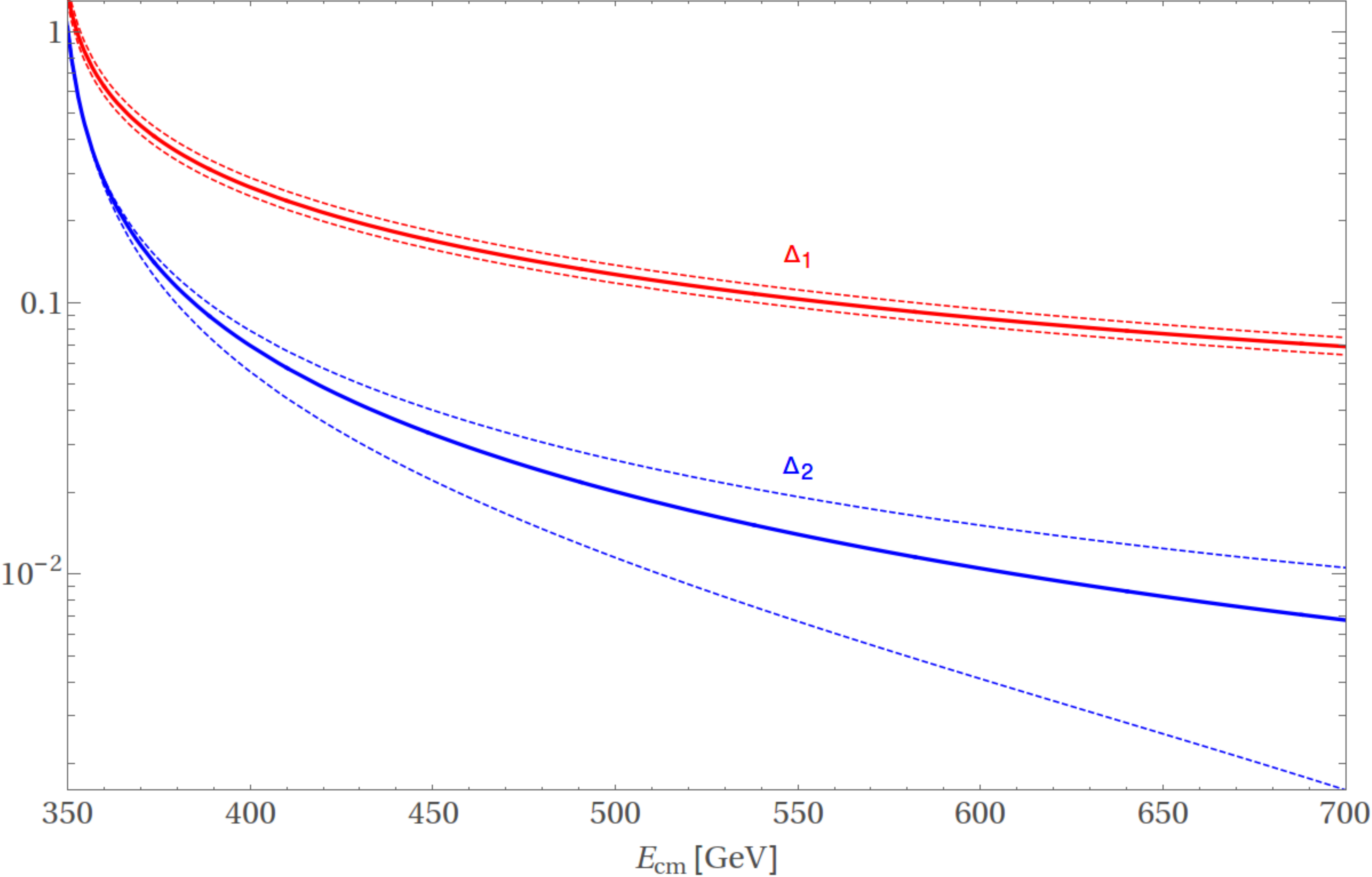}
 \caption{The order $\alpha_s$ and $\alpha_s^2$ corrections $\Delta_1$ and $\Delta_2$ 
 to the LO  $\ttbar$ cross section defined in \eqref{eq:sigDel12} as a function of the c.m. energy.
  The solid lines 
  correspond to the choice $\mu=\sqrt{s}$, the upper and lower dashed lines correspond to 
  $\mu=\sqrt{s}/2$ and $2\sqrt{s}$. }
 \label{fig:deltatt}
 \end{figure}

  One  may represent the $\ttbar$ cross section in the form
  \begin{equation}
  \sigma_{\rm NNLO}= \sigma_{\rm LO} \,(1 + \Delta_1 +  \Delta_2 ) \, .
   \label{eq:sigDel12}
   \end{equation}
   The order $\alpha_s$ and $\alpha_s^2$ corrections $\Delta_1$ and $\Delta_2$ are displayed, for three scales $\mu$, in figure~\ref{fig:deltatt}
    as functions of the c.m. energy $\sqrt{s}$. The changes of $\Delta_1$ and $\Delta_2$ due to scale variations are small -- notice the logarithmic scale 
     on the $y$ axis of figure~\ref{fig:deltatt}.

  We have  included  in the computation of $\sigma_{\rm NNLO}$ also the non-universal  contributions of order $\cO(\as^2)$ (cf. section~\ref{sec:nnlo}) that
  contain the electroweak couplings of quarks $q\neq t$.
   These contributions are, however, very small.  For instance, at $\sqrt{s}$ =  500 GeV they amount to $-0.16\%$ 
    of the total second order correction  $\Delta_2$ 
   defined in \eqref{eq:sigDel12}, and this fraction decreases in magnitude for smaller c.m. energies. 
         
     Our results displayed in figure~\ref{fig:deltatt} agree with the calculation of the $\ttbar$ cross section 
      in \cite{Gao:2014eea}, shown for $\mu=\sqrt{s}$ in figure 1 of this reference. Moreover, considering only the electroweak vector-current contributions to $\sigma_{\rm NNLO}$, 
      we agree with the results of \cite{Dekkers:2014hna,Gao:2014nva}. In addition, we have compared also with the analytically known threshold expansions 
      \cite{Czarnecki:1997vz,Beneke:1997jm,Hoang:1997sj,Bernreuther:2006vp}
      and asymptotic expansions  \cite{Gorishnii:1986pz,Chetyrkin:1996cf,Chetyrkin:1997qi,Chetyrkin:1997pn} of
      $\sigma_{\rm NNLO}$ in the regimes $\alpha_s\ll \beta\ll 1$ (where $\beta$ is the top-quark velocity) and $m_t^2/s\ll 1$, respectively, and find agreement. 
  
      Close to the $\ttbar$ threshold the fixed order perturbative expansion of the cross section and distributions breaks down due to Coulomb singularities.
      (This kinematic regime has been analyzed in detail with effective field  methods.) One can see in  Figure~\ref{fig:sigtt} the onset of the $1/\beta$ singularity in the NNLO cross section for 
       $\sqrt{s}\to 2 m_t$.  We list in table~\ref{tab:delta12} the QCD corrections  $\Delta_1$ and $\Delta_2$ for selected c.m. energies $\sqrt{s}$ for $\mu=\sqrt{s}$.
       With the input values given above  the cross section $\sigma_{\rm NNLO}$ reaches its maximum  at $\sqrt{s}=381.3$ GeV. We obtain $\sigma_{\rm NNLO}(381.3{\rm GeV})=0.843$ pb
        for $\mu=\sqrt{s}$. The numbers in  table~\ref{tab:delta12} and figure~\ref{fig:deltatt} suggest that fixed order 
         perturbation theory can be applied for $\sqrt{s} > 360$ GeV.

 \vspace{2mm}
\begin{table}[tbh!]
\begin{center}
\caption{\label{tab:delta12} The QCD corrections  $\Delta_1$ and $\Delta_2$ defined in \eqref{eq:sigDel12} for several c.m. energies and  $\mu=\sqrt{s}$.}
\vspace{1mm}
\begin{tabular}{|c|cccc|}\hline
  $\sqrt{s}$ [GeV] & 360 & 381.3 & 400  & 500 \\  \hline 
    $\Delta_1$  & 0.627 & 0.352& 0.266 & 0.127 \\
     $\Delta_2$ & 0.281 & 0.110 & 0.070 & 0.020 \\ \hline
\end{tabular}
\end{center}
\end{table}

  Next we turn to differential distributions. We consider the distribution of the cosine of the top-quark 
  scattering angle $\theta_t=\angle(t,e^-)$ in the c.m. frame, the transverse momentum $p_T^t$ of the top quark and 
  of the $\ttbar$ system, $p_T^{\ttbar} = |{\bf k_{T,t}}+ {\bf k_{T,{\bar t}}}|$  with respect to the beam direction, and of the $\ttbar$ invariant mass distribution $M_{\ttbar}$.
  In the following we use the schematic notation LO, NLO, and NNLO for $d\sigma_{\rm LO}/d O$, $d\sigma_{\rm NLO}/d O=(d\sigma_{\rm LO}+d\sigma_1)/d O$, 
  and $d\sigma_{\rm NNLO}/d O=(d\sigma_{\rm LO}+d\sigma_1+ d\sigma_2)/d O$, where $O$
   denotes one of these observables. We confine ourselves to c.m. energies 400 and 500 GeV where the $\ttbar$ cross section is rather large.

\begin{figure}[tbh!]
 \centering
 \includegraphics[width=7.5cm,height=11cm]{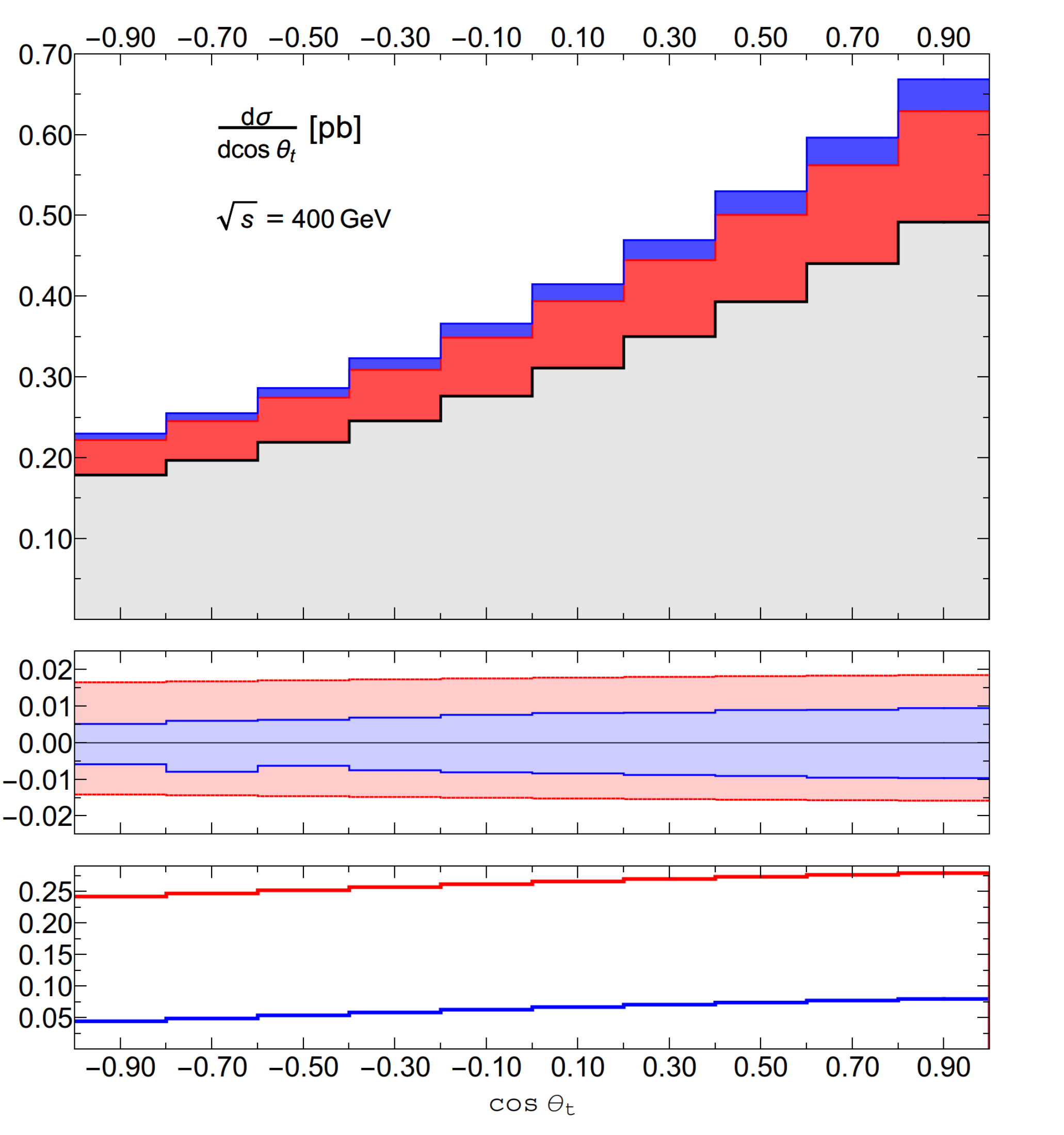}
 \includegraphics[width=7.5cm,height=11cm]{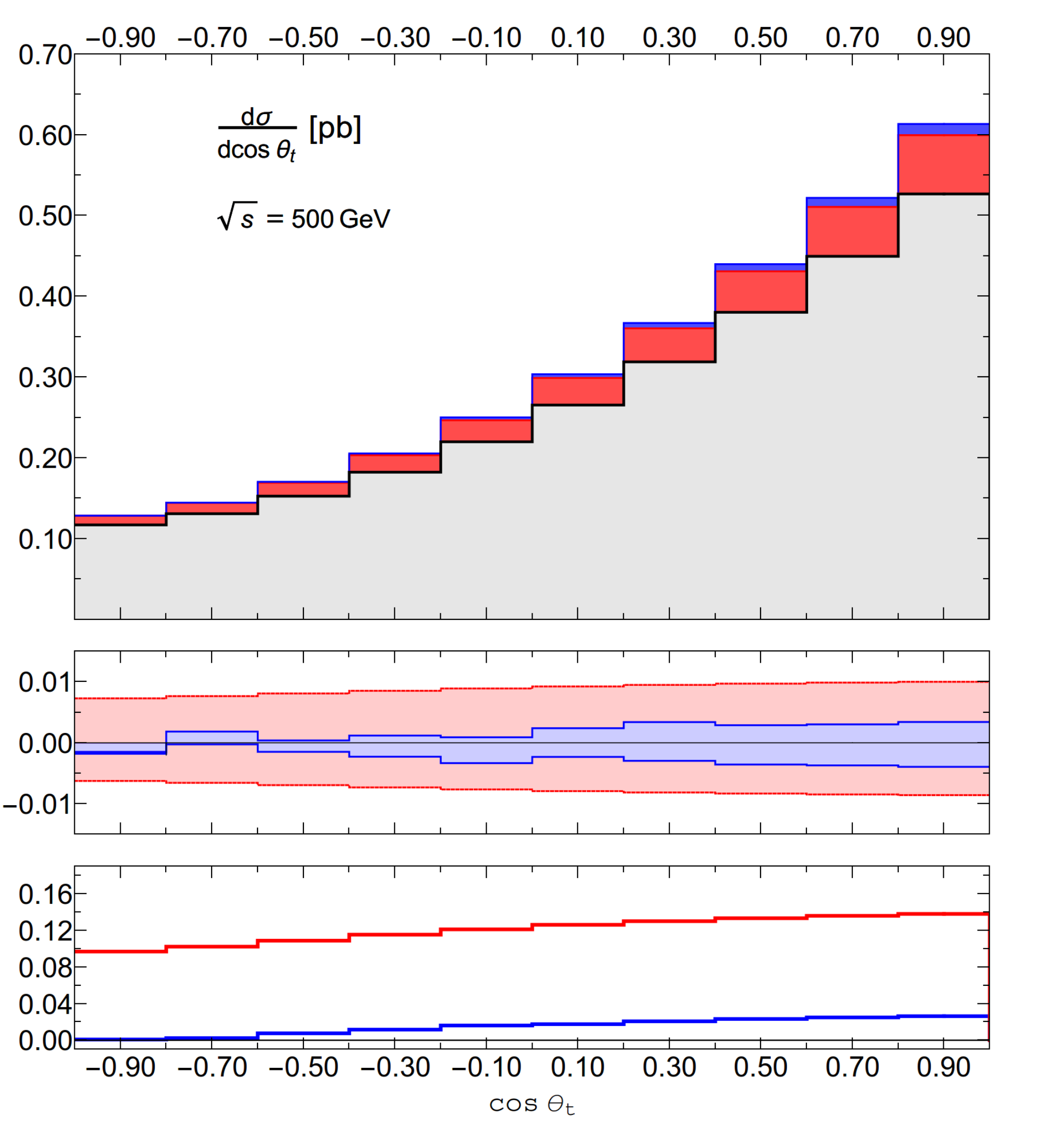}
 \caption{The distribution of $\cos\theta_t$ at $\sqrt{s}= 400$ GeV (plots on the left) and 500 GeV (plots on the right). The upper panels show the distribution
  at LO (grey), NLO (red), and NNLO QCD (blue) for $\mu=\sqrt{s}$. The panels in the middle  show the scale variations ${\rm NLO}(\mu')/{\rm NLO}(\mu=\sqrt{s})-1$ (red band) and 
  ${\rm NNLO}(\mu')/{\rm NNLO}(\mu=\sqrt{s})-1$ (blue band) of the first and second order QCD corrections, where $\sqrt{s}/2\leq\mu'\leq2\sqrt{s}$. The lower panels display
   the ratios ${d\sigma_1}/d\sigma_{\rm LO}$  (red) and ${d\sigma_2}/d\sigma_{\rm LO}$  (blue) for $\mu=\sqrt{s}$. 
   }
 \label{fig:costhet}
 \end{figure}

   The plots in figure~\ref{fig:costhet} display the distribution of $\cos\theta_t$ at $\sqrt{s}= 400$ and 500 GeV at LO, NLO, and NNLO QCD.
   As expected the first- and second-order QCD corrections decrease if one moves further away from threshold. As the panels in the middle of the plots show
    the  inclusion of the order $\alpha_s^2$ correction significantly reduces the dependence of the distribution on variations of the scale. 
    Both the order $\alpha_s$ and order $\alpha_s^2$ follow the same pattern as the leading-order distribution: they are larger in the top-quark forward direction and 
    thus increase the top-quark forward-backward asymmetry, cf. section~\ref{suse:afbtop}. 
    The ratios ${d\sigma_1}/d\sigma_{\rm LO}$  and  ${d\sigma_2}/d\sigma_{\rm LO}$  for $\sqrt{s}=400$ GeV and $\mu=\sqrt{s}$ shown in the lower panel of the left 
    plot in figure~\ref{fig:costhet} agree with the corresponding result given  in \cite{Gao:2014eea}.

\begin{figure}[tbh!]
 \centering
 \includegraphics[width=7.5cm,height=11cm]{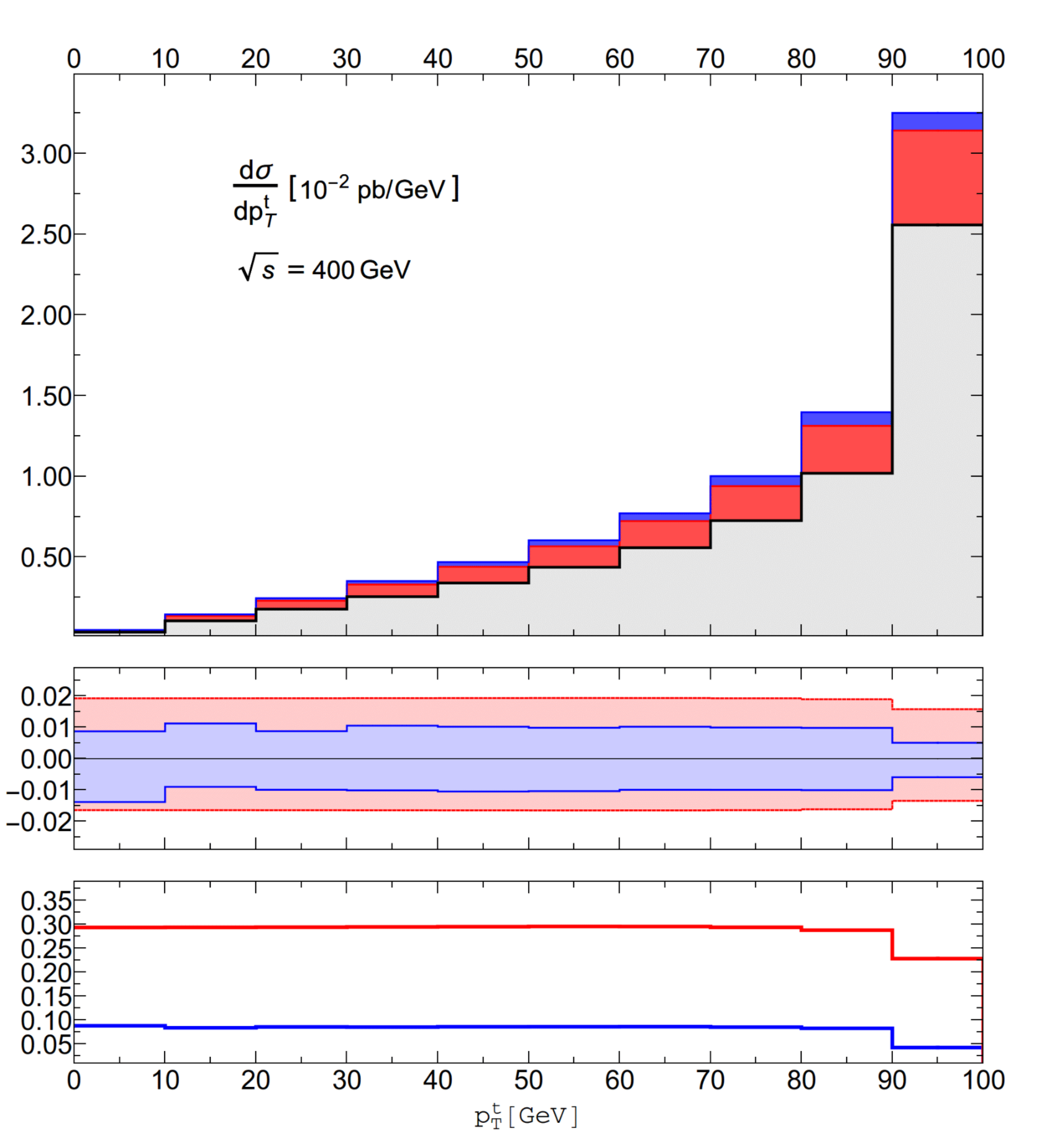}
 \includegraphics[width=7.5cm,height=11cm]{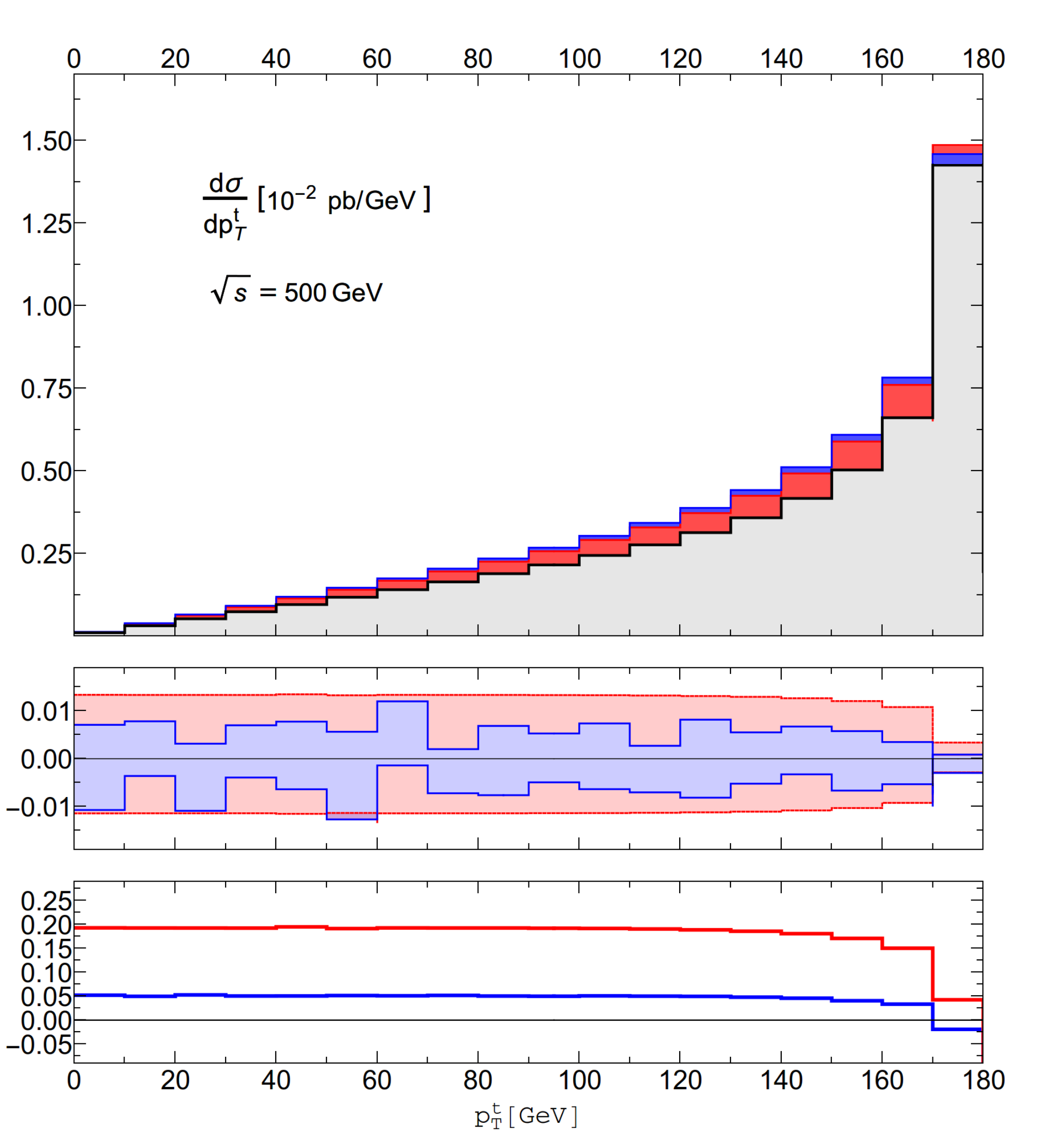}
 \caption{The distribution of the  transverse  momentum $p_T^t$ of the top quark at $\sqrt{s}= 400$ GeV (plots on the left) and 500 GeV (plots on the right). 
 The meaning of the upper, middle, and lower panels is as in figure~\ref{fig:costhet}. }
 \label{fig:PTt} 
 \end{figure}

 Figure~\ref{fig:PTt} displays the distribution of the top-quark  transverse  momentum $p_T^t$ at  $\sqrt{s}= 400$ GeV  and 500 GeV  at LO, NLO, and NNLO QCD.
  Also here the order $\alpha_s$ and $\alpha_s^2$ corrections become smaller when the c.m. energy is increased from  400  to 500 GeV, but the $\cO(\alpha_s^2)$ correction is 
  still $\sim 5\%$ for most of the $p_T^t$ bins. The scale variation of the distribution becomes smaller at NNLO for all $p_T^t$ bins. 
   The maximum top-quark  transverse  momentum is 99.77 GeV (180.15 GeV) at the  c.m. energy 400 GeV (500 GeV) for a top-quark mass $m_t=173.34$ GeV that we use.
    The ratios  ${d\sigma_1}/d\sigma_{\rm LO}$  and  ${d\sigma_2}/d\sigma_{\rm LO}$     
     for $\sqrt{s}=400$ GeV and $\mu=\sqrt{s}$ shown in the lower panel of the left 
    plot in figure~\ref{fig:PTt} agree with the corresponding plot displayed  in \cite{Gao:2014eea}, except for the last bin at $(p_T^t)_{\rm max}$.
    This is apparently due to the fact that  a slightly  lower value of the top-quark mass  was used in \cite{Gao:2014eea}. This shifts  $(p_T^t)_{\rm max}$ at 400 GeV 
    to a value slightly above 100 GeV.

\begin{figure}[tbh!]
 \centering
 \includegraphics[width=7.5cm,height=11cm]{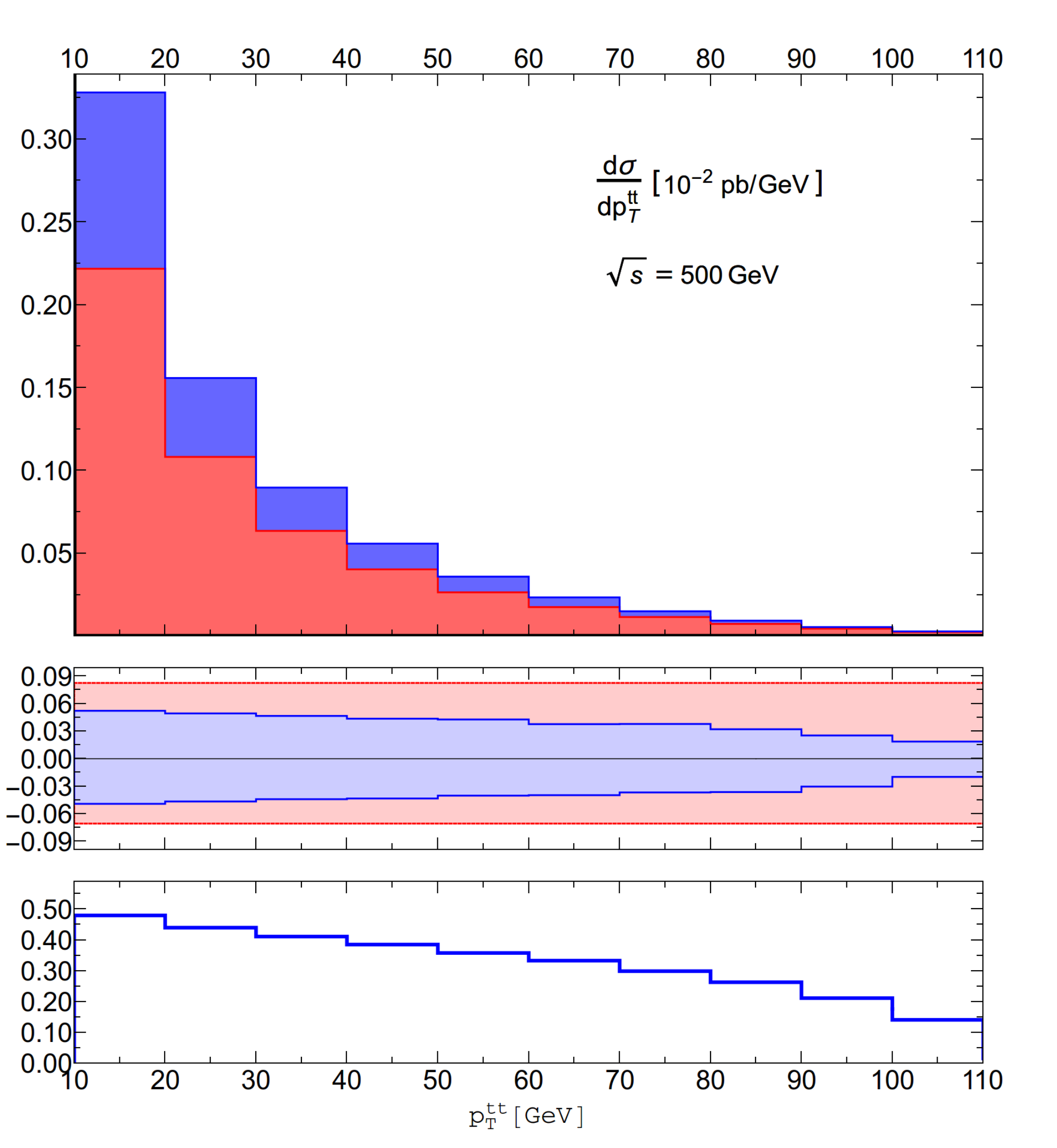}
 \includegraphics[width=7.5cm,height=11cm]{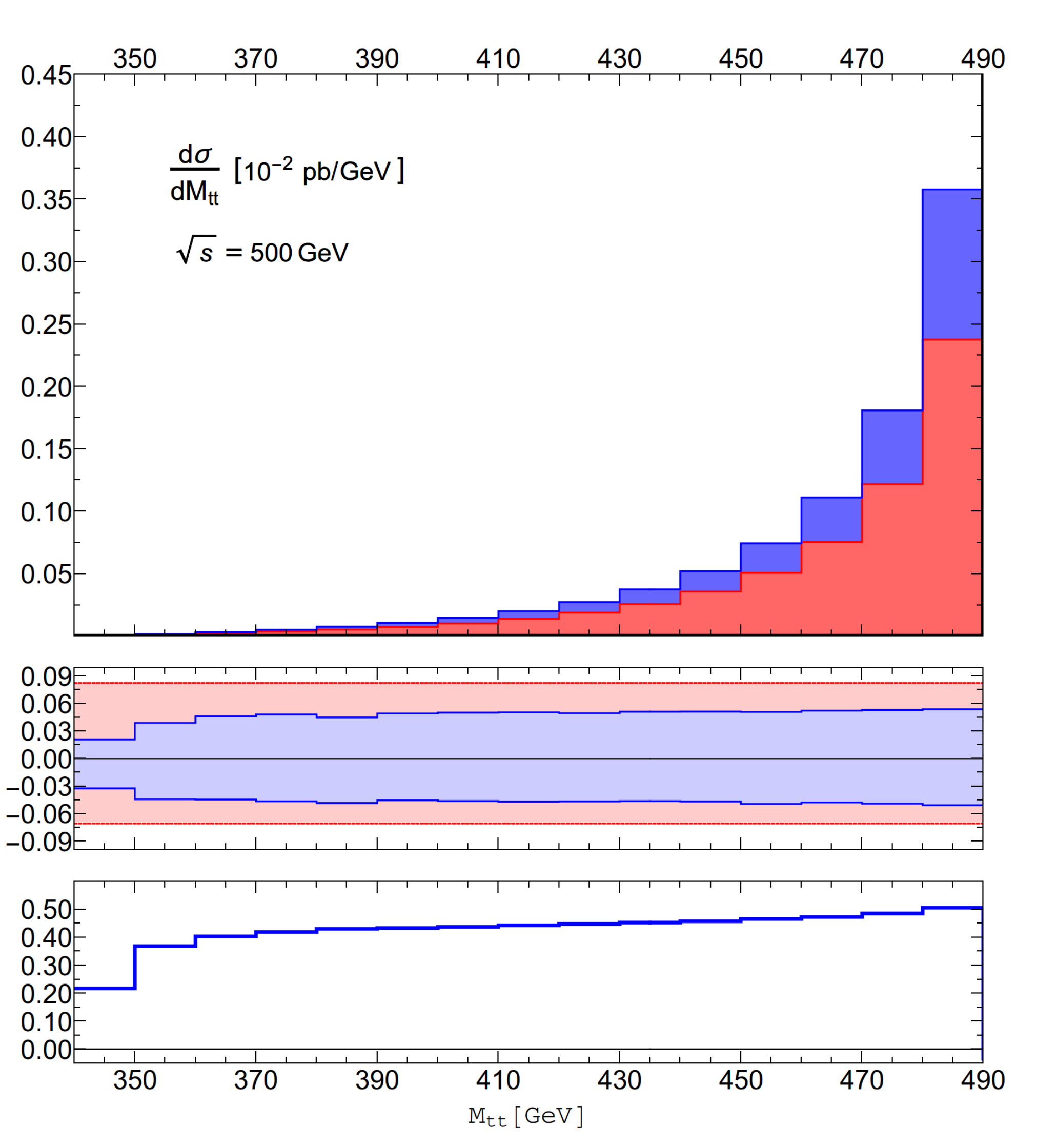}
 \caption{The distribution of the transverse  momentum $p_T^{\ttbar}$ of the $\ttbar$ system for events with $p_T^{\ttbar}\geq 10$ GeV (plots on the left),
 and the  $\ttbar$ invariant mass distribution for events with $M_{\ttbar}\leq 490$ GeV (plots on the right) at $\sqrt{s}= 500$ GeV for $\mu=\sqrt{s}$. 
  The upper panels show the  distributions at NLO (red) and NNLO QCD (blue).
 The panels in the middle  show the scale variations ${\rm NLO}(\mu')/{\rm NLO}(\mu=\sqrt{s})-1$ (red band) and 
  ${\rm NNLO}(\mu')/{\rm NNLO}(\mu=\sqrt{s})-1$ (blue band) of the first and second order QCD corrections, where $\sqrt{s}/2\leq\mu'\leq2\sqrt{s}$. The lower panels display
   the ratio $d\sigma_2/d\sigma_1$ for $\mu=\sqrt{s}$.  }
 \label{fig:PTTt}
 \end{figure}

 The left plots of figure~\ref{fig:PTTt} show, for  $\sqrt{s}= 500$ GeV, the distribution of the transverse momentum of the $\ttbar$ system, $p_T^{\ttbar}$, for events with 
 $p_T^{\ttbar}\geq 10$ GeV. The $p_T^{\ttbar}$ cut removes the LO QCD contribution and events with very soft massless parton radiation at order $\alpha_s$ and $\alpha_s^2$.
    For $\sqrt{s}= 500$ GeV the maximum $p_T^{\ttbar}$ is 129.81 GeV, but events with $p_T^{\ttbar}$ near this value are very rare. The NLO and NNLO QCD corrections increase 
    significantly towards small $p_T^{\ttbar}$. This is due to logarithmic enhancement in the variable $p_T^{\ttbar}$ that arises in the sum of the order $\as^2$ three-parton and 
    four-parton contributions. In the bin 
    $10~{\rm GeV}\leq p_T^{\ttbar}\leq 20~{\rm GeV}$ the order $\alpha_s^2$ correction is almost $50\%$ of the NLO correction. The fixed-order calculation  
    of the distribution becomes unreliable for small $p_T^{\ttbar}$; the logarithms should be resummed. But this is beyond the scope of this paper.
    An analogous statement applies to the right plots of figure~\ref{fig:PTTt} that show the $\ttbar$ invariant mass distribution for events with  $M_{\ttbar}\leq 490$ GeV.
    This cut removes  the LO QCD contribution and events with very soft parton radiation.

\subsection{The forward-backward asymmetry}
\label{suse:afbtop}

The top-quark forward-backward asymmetry  $\afb$  is
 defined as the number of $t$ quarks  observed in the forward hemisphere
minus the number of $t$ quarks in the backward hemisphere, divided by
the total number of observed $t$  quarks. 
\begin{align}
\afb = \frac{N_F - N_B}{N_F + N_B} \, .
\label{def0afb} 
\end{align}
Forward and backward
hemispheres are defined with respect to a certain IR-safe axis. 
For top-quark production in $e^-e^+$ collisions, the 
top-quark direction of flight is a good choice, because this direction can be reconstructed for instance
 with lepton plus jets events from $\ttbar$ decay.
 This axis is  infrared- and collinear-safe for massive quarks. Thus $\afb$ is computable
in perturbation theory.
 As long as we consider  $\afb$ below the four-top threshold, i.e., as long as we do not include the 
 $t{\bar t} t{\bar t}$ final state in the computation of the forward-backward asymmetry,  
 $\afb$ can be expressed in terms of the antisymmetric and symmetric $\ttbar$ cross section 
\begin{align}
\afb = \frac{\sia}{\sis}
\label{defafb},
\end{align}
where the antisymmetric cross section $\sigma_A$ is defined by
\begin{align}
\sia = & \int_0^1\frac{d\sigma}{d\cos\theta_t} d\cos\theta_t - \int_{-1}^0\frac{d\sigma}{d\cos\theta_t} d\cos\theta_t \, ,
\label{defsia}
\end{align}
and  $\sis$ is equal to the cross section calculated
 in section~\ref{suse:xsecdist}.
As in the previous subsection $\theta_t$ denotes the angle between the incoming electron and the top-quark direction of flight 
 in the $e^- e^+$ c.m. frame. 
   The NLO QCD and electroweak corrections to $\afb$ for massive quarks were determined in \cite{Jersak:1979uv,Arbuzov:1991pr,Djouadi:1994wt}
   and  \cite{Beenakker:1991ca,Bardin:1999yd}, respectively. The order $\alpha_s^2$ corrections to $\afb$ were calculated in the limit of massless quarks in 
   \cite{Altarelli:1992fs,Ravindran:1998jw,Catani:1999nf,Weinzierl:2006yt} and for top quarks with full mass dependence in \cite{Waning2011,Gao:2014eea}.
 
 In order to compute \eqref{defafb} to second order in $\alpha_s$
  in the spirit of perturbation theory, we Taylor-expand
  \eqref{defafb} to second order in $\alpha_s$ and obtain
 
 \begin{equation}
  \afb^{\rm NNLO} = \afb^{\rm LO} (1 + A_1 + A_2) \, ,
  \label{eq:ANNLOex}
 \end{equation}
where $\afb^{\rm LO}$ is
the forward-backward asymmetry at Born level, and  $A_{1}$ and $A_{2}$
are the QCD corrections of $\cO(\as)$ and  $\cO(\as^2)$, respectively. These terms are given by
\begin{align}
 \afb^{\rm LO} =& \frac{\sia^{(2,0)}}{\sis^{(2,0)}} \, ,\label{eq:afb0}\\
A_{1} =& \sum\limits_{i=2,3} \big[\frac{\sia^{(i,1)}}{\sia^{(2,0)}} \;-\;\frac{\sis^{(i,1)}}{\sis^{(2,0)}} \big] \, ,\label{eq: afb1} \\
A_{2} =& \sum\limits_{i=2,3,4} \big[ \frac{\sia^{(i,2)}}{\sia^{(2,0)}} \;-\;
\frac{\sis^{(i,2)}}{\sis^{(2,0)}} \big]
 - \frac{\sis^{(2,1)}+\sis^{(3,1)}}{\sis^{(2,0)}} \, A_1  \, .\label{eq:afb2}
\end{align}
 In analogy to the  notation in \eqref{eq:strucsig}
 the first number $i$ in the superscript $(i,j)$ labels the number of partons 
  in the final state and the second one the order of $\as$.

 Table~\ref{tab:afbex} contains our results for the top-quark forward-backward 
 asymmetry using the expansion \eqref{eq:ANNLOex} -- \eqref{eq:afb2}  
 for several c.m. energies and for the input values as given at the beginning of section~\ref{sec:restt}. Notice that $\afb^{\rm NLO} =\afb^{\rm LO} (1 + A_1).$
  The central values refer to the scale choice $\mu=\sqrt{s}$ and the given uncertainties are 
  obtained by varying $\mu$ between $\sqrt{s}/2$ and $2\sqrt{s}.$
  We have included in $A_2$ also the non-universal  contributions  that
  contain the electroweak couplings of quarks $q\neq t$. (We remark that the square of the diagrams where $\gamma^*/Z^*$ couple to $q\neq t$ 
   and the $\ttbar$ pair is produced by gluon splitting does not contribute to the antisymmetric cross section.)
   These contributions are, however, small. The ratio $r=A_2^{\rm non}/A_2$ of the non-universal and the total order $\as^2$ correction 
    increases with increasing c.m. energy. We have $r= -0.16\%, -1\%,$ and $-2.4\%$ for $\sqrt{s}=400~{\rm GeV},500~{\rm GeV},$ and  $700~{\rm GeV},$ respectively.
   
    As one can see from table~\ref{tab:afbex}, close to the $\ttbar$ threshold, at $\sqrt{s} \simeq 360$ GeV, 
   fixed order perturbation theory is  no longer reliable because the second order correction $A_2$ to the forward-backward asymmetry is larger than 
    the first order correction $A_1$. For $\sqrt{s}> 380$ GeV, the ratio $|A_2/A_1|$ becomes smaller than one. Notice that 
     the order $\alpha_s^2$ correction is significant as compared to  the first order one: $|A_2/A_1| \gtrsim 0.5$
     for the  c.m. energies listed in table~\ref{tab:afbex}. The uncertainties of $\afb^{\rm NNLO}$ due to scale variations are small, but these uncertainties do not decrease,
       in the range of c.m. energies  given in table~\ref{tab:afbex}, compared to the scale uncertainties of $\afb^{\rm NLO}$.
       The numbers $\delta\afb^{\rm NNLO}$ in the last column of table~\ref{tab:afbex} signify the change of $\afb^{\rm NNLO}$ if our input value for the top-quark mass 
       is changed by $\pm0.5$ GeV. For a given c.m. energy $\afb^{\rm NNLO}$ increases if the top-quark mass is decreased and vice versa. One expects that the top-quark mass can be 
       measured with a much smaller uncertainty than  $\pm 0.5$ GeV from a $\ttbar$ threshold scan at a future $e^-e^+$ collider \cite{Moortgat-Picka:2015yla,Vos:2016til}. 
     
     The two-parton, i.e., the $\ttbar$ contribution to $\afb$ is separately IR-finite, both at order $\alpha_s$ and at order $\alpha_s^2$ \cite{Bernreuther:2006vp}.
      In the range of c.m. energies  given in table~\ref{tab:afbex}, the $\ttbar$ final state makes the largest contribution both to $A_1$ and $A_2$. For $\sqrt{s}\lesssim 500$ GeV
      it is significantly larger than the respective contribution from the three- and four-parton final states. 
      Here, we have computed the  $\ttbar$ contribution to $A_1$ and $A_2$   with the antenna-subtracted two-parton matrix elements of sections~\ref{sec:lonlo} and~\ref{sec:nnlo}, while 
       it was computed in \cite{Bernreuther:2006vp} with the unsubtracted $\ttbar$ matrix elements. We agree with the results of \cite{Bernreuther:2006vp}.
       This serves as a check of our calculation. 
      The sum of the three- and four-parton contribution $A^{(3)}_2+A^{(4)}_2$ to $A_2$  is also IR-finite. 
       It was computed in \cite{Waning2011} with an NLO subtraction scheme, namely dipole subtraction with massive quarks \cite{Catani:2002hc}. 
        We agree with the results of \cite{Waning2011}.

\vspace{2mm}
\begin{table}[tbh!]
\begin{center}
\caption{\label{tab:afbex} The top-quark forward-backward asymmetry  at LO, NLO, and NNLO QCD for several c.m. energies
 using the expansion  \eqref{eq:ANNLOex}. The numbers are given in percent.}
\vspace{1mm}
\begin{tabular}{|c|ccc|cc|c|}\hline
  $\sqrt{s}$ [GeV] & $\afb^{\rm LO} $ [\%]  & $\afb^{\rm NLO}$ [\%] & $\afb^{\rm NNLO}$ [\%] & $A_1$ [\%] & $A_2$  [\%] & $\delta\afb^{\rm NNLO}$ [\%]\\  \hline 
    360 & 14.94 & $15.54^{+0.05}_{-0.04}$ & $16.23^{+0.12}_{-0.10}$ & $4.01^{+0.35}_{-0.29}$ &  $4.58^{+0.46}_{-0.38}$ & $\pm  0.59  $ \\[.3em]
    400 & 28.02 & $28.97^{+0.08}_{-0.07}$ & $29.63^{+0.11}_{-0.10}$ & $3.41^{+0.29}_{-0.25}$ &  $2.36^{+0.11}_{-0.11}$ & $\pm  0.27 $ \\[.3em]
    500 & 41.48 & $42.42^{+0.08}_{-0.07}$ & $42.91^{+0.08}_{-0.07}$ & $2.28^{+0.19}_{-0.16}$ &  $1.18^{+0.01}_{-0.01}$ & $\pm  0.13 $ \\[.3em] 
    700 & 51.34 & $51.81^{+0.04}_{-0.03}$ & $52.05^{+0.04}_{-0.04}$ & $0.91^{+0.07}_{-0.06}$ &  $0.47^{+0.01}_{-0.01}$ & $\pm  0.06 $ \\[.3em] \hline
\end{tabular}
\end{center}
\end{table}

 Measurements of forward-backward asymmetries or simulations with Monte Carlo event generators correspond to computations where the 
 ratio in \eqref{defafb} is not Taylor-expanded. Using our results for $\sia$ and $\sis$ at  $\cO(\as)$ and  $\cO(\as^2)$, respectively, and
 the input values as given at the beginning of section~\ref{sec:restt}, we give in table~\ref{tab:afbunex} the values for the unexpanded version of the forward-backward 
  asymmetry at NLO and NNLO QCD. 
 (For ease of notation, we use the same symbols as in
  Table~\ref{tab:afbex}.) The central values and the uncertainties refer again to the scales $\mu=\sqrt{s}$ and $\mu=\sqrt{s}/2$ and $2\sqrt{s}$, respectively.
  We vary $\mu$ simultaneously in the numerator and denominator of \eqref{defafb}.

  The top-quark forward-backward asymmetry at NNLO QCD was computed before in ref.~\cite{Gao:2014eea} in the unexpanded version with values of $m_t$ and $\alpha_s$
   that differ slightly from the ones that we use here. Our results of table~\ref{tab:afbunex} agree with those given in table~1 of that reference.

\vspace{2mm}
\begin{table}[tbh!]
\begin{center}
\caption{\label{tab:afbunex} The unexpanded version of the top-quark forward-backward asymmetry \eqref{defafb} for several c.m. energies.
 The numbers are given in percent.}
\vspace{1mm}
\begin{tabular}{|c|ccc|}\hline
  $\sqrt{s}$ [GeV] & $\afb^{\rm LO} $  [\%]     & $\afb^{\rm NLO}$ [\%] & $\afb^{\rm NNLO}$ [\%] \\  \hline 
    360            & 14.94 & $15.31^{+0.02}_{-0.02}$  & $15.82^{+0.08}_{-0.06}$  \\[.3em]
    400            & 28.02 & $28.77^{+0.05}_{-0.04}$  & $29.42^{+0.10}_{-0.09}$  \\[.3em]
    500            & 41.48 & $42.32^{+0.06}_{-0.05}$  & $42.83^{+0.08}_{-0.07}$  \\[.3em] 
    700            &  51.34 &  $51.78^{+0.03}_{-0.03}$ & $52.03^{+0.04}_{-0.04}$ \\[.3em] \hline          
\end{tabular}
\end{center}
\end{table}

 One may take the spread between the values of the expanded and unexpanded versions of $A_{FB}^{\rm NLO}$  and  $A_{FB}^{\rm NNLO}$ given in 
  tables~\ref{tab:afbex},~\ref{tab:afbunex}  as an estimate of the uncalculated higher order corrections.
  Alternatively, one may add
  the scale uncertainties and the uncertainties due to $\delta m_t=\pm 0.5$ GeV of the expanded version (cf. table~\ref{tab:afbex}) linearly
  and take this residual uncertainty of our  prediction of $\afb^{\rm NNLO}$. This yields an uncertainty of  $0.4 \%$ and $0.2\%$ at $\sqrt{s}=400$ and $500$ GeV, respectively.
  This uncertainty is in accord with the spread between the expanded and unexpanded  results listed in tables~\ref{tab:afbex} and~\ref{tab:afbunex}.   
 This uncertainty is significantly smaller than the projected experimental precision of top-quark $\afb$ measurements at future 
 electron-positron colliders \cite{Devetak:2010na,Amjad:2015mma}. This observable has a high sensitivity to precisely determine the neutral current couplings 
 of the top quark and probe for anomalous couplings \cite{Devetak:2010na,Rontsch:2015una,Khiem:2015ofa,Janot:2015yza,Amjad:2015mma}.

\section{Summary}
 \label{sec:sumconc}

We have formulated, within the antenna subtraction framework, the set-up for calculating the  production of a massive quark-antiquark pair in electron-positron 
collisions at NNLO QCD. Our approach is fully differential in the phase-space variables and can be used to compute any infrared-safe observable.
 We have applied this formalism to $\ttbar$ production in the continuum and we have calculated, besides the $\ttbar$ cross section also several distributions in order to 
 signify the usefulness of this approach, namely the $\cos\theta_t$ and transverse momentum distribution of the top quark, the transverse momentum of the $\ttbar$ system
 and the $\ttbar$ invariant mass distribution. The NNLO QCD corrections are sizable for c.m. energies  not too far away from the $\ttbar$ threshold. We have also computed the
 top-quark forward-backward asymmetry, which is an important observable for determining the neutral-current couplings of the top quark at future lepton colliders, at NNLO QCD.
  Our result agrees with previous calculations \cite{Waning2011,Gao:2014eea} of this asymmetry at order $\alpha_s^2$.
  
  Our set-up may be used to investigate a number of other reactions at NNLO QCD where a massive quark-pair is produced by an uncolored initial state.
  Of interest for future lepton colliders would be the production of $\ttbar$ pairs with spin correlations included.
  Other applications include the production of charm and bottom quarks, in particular at the $Z$-boson resonance.

 \pagebreak
\appendix
\section{Phase-space mappings that involve massive particles}
\label{sec:AppA}
 We describe here the phase-space mappings that are used in the construction of the antenna subtraction terms
 of section~\ref{sec:lonlo} and~\ref{sec:nnlo}. The momentum mappings required in our case are related either to 
 a single or double unresolved parton configuration in the final state. These mappings must  obey four-momentum conservation, must keep the 
  mapped momenta on their respective mass shell, and the mapped momenta must converge to the correct momentum configurations in the  the soft and 
  collinear limits.  We follow the mapping procedures of \cite{GehrmannDeRidder:2007jk}, which apply to the case where all 
  partons are massless, respectively of \cite{Abelof:2011ap,AbelGerh2016}
  where  the massless case was extended to configurations involving massive partons. The analytic formulas  of 
  Abelof and Gehrmann-De Ridder that keep the mapped momenta on-shell in the massive case have not been published so far \cite{AbelGerh2016}.
  Therefore we describe below an alternative mapping method for computing the observables used in this paper. 

\subsection{Three parton final states}
\label{suse:map3p}
We consider the final state $Q(k_1){\bar Q}(k_2)g(k_3)$. The NLO subtraction term $d\sigma_{Q\bar{Q}g}^S$ of
 eq.~\eqref{sub::NLO} and the NNLO subtraction terms 
 $d \sigma^{T,b,Q \bar{Q} g}_\mss{NNLO}$   and    
 $d \sigma^{T,c,Q \bar{Q} g}_\mss{NNLO}$ of eq.~\eqref{QQg::sub::b} and \eqref{QQg::sub::c::1} depend on mapped momenta
  obtained from a $3\to 2$ mapping $k_1, k_3, k_2\to \widetilde{ k_{13} }, \widetilde{k_{32}}$.
  Let's consider the mapping $k_1, k_3, k_2\to p_I\equiv \widetilde{ p_{13} }, p_J\equiv \widetilde{p_{32}}$ defined in 
  \cite{GehrmannDeRidder:2007jk} and in appendix B1.1 of \cite{Abelof:2011ap}:
  \begin{flalign} 
   p_I & =  x k_1^\mu + r k_3^\mu+ z k_2^\mu \, , \nonumber \\
   p_J & =  (1-x) k_1^\mu + (1-r) k_3^\mu+ (1-z) k_2^\mu \, ,
   \label{eq:p323p}
  \end{flalign}
 where the parameters $x,r,z$ are given in \cite{GehrmannDeRidder:2007jk,Abelof:2011ap}.
 The mapping \eqref{eq:p323p} satisfies four-momentum conservation, $p_I+p_J=k_1+k_3+k_2$,
  and the mapped momenta behave correctly when the gluon becomes soft:
  $p_I\to k_1, p_J\to k_2$ if $k_3\to 0$. If all three partons were massless, the mapped momenta 
  remain massless,  $p_I^2=p^2_J= 0$ \cite{GehrmannDeRidder:2007jk}. However, for a massive quark  $Q$ 
  modified formulas  must be used for the  parameters $x,r,z$ in order to get 
  $p_I^2,p^2_J = m_Q^2$  \cite{AbelGerh2016}. We recall that on-shellness of the mapped momenta is crucial
   for deriving the correct integrated antenna subtraction terms from the unintegrated ones.
   
 Here we describe, as an alternative to the analytic formulas of  Abelof and Gehrmann-De Ridder \cite{AbelGerh2016}, a numerical method to 
 obtain on-shell mapped momenta $\widetilde{ k_{13} }, \widetilde{k_{32}}$. We use the mapping  \eqref{eq:p323p}
 with the parameters $x,r,z$ given in \cite{Abelof:2011ap} for an intermediate step. Four-momentum conservation in the 
 $e^-e^+$ c.m. frame reads:
 \begin{equation} \label{eq:4momc}
   \sqrt{s} = p_{0I} +  p_{0J} \, , \quad 0 = {\bf p}_{I} +  {\bf p}_{J} \, .
 \end{equation}
 The second equation is the crucial one. It allows to rescale the 3-momenta 
 by a factor $\xi$ such that the
   4-momenta $p^\mu_I, p^\mu_J$ are transformed into on-shell 4-momenta $k_I, k_J$
   with mass $m_I=m_J=m_Q$ without destroying 4-momentum conservation.
   \begin{equation} \label{eq:scaletr3p} 
 {\bf k}_i = \xi {\bf p}_i \, , \quad k_{0i} = \sqrt{m_i^2 + \xi^2({p}_{0i}^2 - {p}^2_i)} \, , \qquad i = I,J \, ,
 \end{equation}
 where $\xi$ is the solution of the equation 
 \begin{equation}
  \sqrt{s} = \sqrt{m_I^2 + \xi^2({p}_{0I}^2 - {p}^2_I)} + \sqrt{m_J^2 + \xi^2({p}_{0J}^2 - {p}^2_J)} \, .
   \label{eq:xisol3p} 
  \end{equation}
  Eq.~\eqref{eq:xisol3p} can be solved numerically by iteration using the Newton-Raphson method. 
  One can start the iterative solution of  \eqref{eq:xisol3p}  with the
 value $\xi =\sqrt{(1-(2m_Q/\sqrt{s})^2)}.$ We found that a few iterations $(n\leq 6)$ are enough 
 to get an accuracy of $10^{-14}\sqrt{s}/[{\rm GeV}]$. 
 Eqs.~\eqref{eq:scaletr3p} and \eqref{eq:xisol3p} are completely analogous to the procedure used in the phase-space generator
 {\tt RAMBO}  \cite{Kleiss:1985gy} for constructing massive four-momenta from massless ones.

\subsection{Four parton final states}
\label{suse:map4p}
We consider the final states $Q(k_1){\bar Q}(k_2)a(k_3)b(k_4)$, where $ab =q{\bar q}, gg$.
For the a- and b-type subtraction terms of section~\ref{suse:RR} we need $3\to 2$ and $4\to 2$ momentum mappings associated with single and double unresolved 
configurations.

\subsubsection*{$3\to 2$ mappings for a-type antenna subtraction terms:}
 For evaluating the  a-type subtraction term 
$d\sigma^{S,a,Q \bar{Q} q \bar{q}}_\mss{NNLO}$  of \eqref{eq:SQQqqa}
 one needs mapped momenta obtained by the two $3\to 2$ mappings
 \begin{equation} \label{eq:32a1}
  k_1, k_3, k_4 \to \widetilde{k_{13}}, \widetilde{k_{34}} \, , \qquad k_3, k_4, k_2 \to\widetilde{k_{34}}, \widetilde{k_{42}} \, ,
 \end{equation}
 where the massless (anti)quark with momentum $k_3$, respectively $k_4$ becomes unresolved. For definiteness we describe for the mapping on the 
 left side of \eqref{eq:32a1} how one can obtain, with a procedure analogous to  that of section~\ref{suse:map3p},
  mapped momenta that satisfy four-momentum conservation and the on-shell conditions $\widetilde{k^2_{13}}=m_Q^2$, $\widetilde{k^2_{34}}=0$.
  We use again in an intermediate step the $3\to 2$ mapping 
   \begin{flalign} 
   p_I & =  x k_1^\mu + r k_3^\mu+ z k_4^\mu \, , \nonumber \\
   p_J & =  (1-x) k_1^\mu + (1-r) k_3^\mu+ (1-z) k_4^\mu \, ,
   \label{eq:p423p}
  \end{flalign}
 where the parameters $x,r,z$ are given in appendix B1.1 of \cite{Abelof:2011ap}. We have $p^2_I \neq m_Q^2$ and $p_J^2 \neq 0$ for general configurations $k_j$.
  In the case of four-parton final states, momentum conservation in the $e^-e^+$ c.m. frame reads in terms of the mapped momenta $p_I, p_J$:
 \begin{equation} \label{eq:4momee}
    Q_2^\mu =  p^\mu_I + p^\mu_J \, ,
 \end{equation}
 where $ Q_2^\mu =(\sqrt{s}-k_{02}, -{\bf k}_2)$. 
 Now we boost to the rest frame IS' of  $Q_2^\mu$ with the boost vector ${\boldsymbol\beta_2} = -{\bf k_2}/Q_{02}$.
  Four-momentum conservation in IS' reads
  \begin{equation} \label{eq:4momrest}
  Q_{02}' = p'_{0I} + p'_{0J} \, , \qquad 0 = {\bf p'}_{I} + {\bf p'}_{J} \, .
  \end{equation}
  As above, the second equation allows to rescale the 3-momenta by a factor $\xi$ such that the
   4-momenta $p'^\mu_I, p'^\mu_J$ are transformed into on-shell 4-momenta $k'^\mu_I, k'^\mu_J$ with mass $m_I=m_Q$ and $m_J=0$, respectively,
   without destroying 4-momentum conservation. 
   \begin{equation} \label{eq:scaletrbo} 
 {\bf k'}_i = \xi {\bf p'}_i \, , \quad k'_{0i} = \sqrt{m_i^2 + \xi^2({p'}_{0i}^2 - {p'}^2_i)} \, , \qquad i = I,J \, ,
 \end{equation}
 where $\xi$ is the solution of the equation 
 \begin{equation} \label{eq:xisolbo} 
  Q_{02}' = \sqrt{m_I^2 + \xi^2({p'}_{0I}^2 - {p'}^2_I)} + \sqrt{m_J^2 + \xi^2({p'}_{0J}^2 - {p'}^2_J)} \, .
  \end{equation}
Again, eq.~\eqref{eq:xisolbo} can be solved numerically by iteration using the Newton-Raphson method. In this case it is advantageous to start the iteration with the
 value $\xi =\sqrt{(1-(m_Q/Q_{02}')^2)}.$ 
 A few iterations $(n\leq 6)$ are enough to get an accuracy of $10^{-14}Q_{02}'/[{\rm GeV}]$. \\
 Finally we boost  $k'^\mu_I, k'^\mu_J$ back to the $e^-e^+$ c.m. frame IS with the boost vector  $-{\boldsymbol\beta_2}$ 
  and we obtain $k_I \equiv \widetilde{k_{13}}, k_J\equiv \widetilde{k_{34}}$. 
  The on-shell 4-momenta $k^\mu_I, k^\mu_J$ satisfy 4-momentum conservation in IS and behave  correctly in all singular limits. 
  
  For the second set of momenta in  \eqref{eq:32a1} mapped on-shell momenta are constructed in completely analogous fashion with the 
    `spectator' $k_2$ replaced by $k_1$. 
  The above procedure applies  also to the  various $3\to 2$ mappings that  are required  for the a-type subtraction term 
  $d\sigma^{S,a, Q \bar{Q} g g}_\mss{NNLO}$   of \eqref{sub:gga}. 
  
  Abelof and Gehrmann-De Ridder have derived analytic formulas for the  mapped on-shell 4-momenta $k^\mu_I, k^\mu_J$  \cite{AbelGerh2016}.
  
\subsubsection*{$4\to 2$ mappings for b-type antenna subtraction terms:}
 The b-type antenna subtraction term $d \sigma^{S,b,2, Q \bar{Q} q \bar{q}}_\mss{NNLO}$ of \eqref{eq:Sb2qqbar}
 is evaluated with mapped momenta that are obtained by a  $4\to 2$ mapping
\begin{equation}\label{eq:4to2}
 k_1, k_3, k_4, k_2 \to k_{I} \equiv \widetilde{ k_{134}}, \;   k_{J} \equiv \widetilde{ k_{342}} \, .
\end{equation}
As an intermediate step we use the $4\to 2$ mapping $k_1, k_3, k_4, k_2\to p_I, p_J$, where
  \begin{flalign} 
   p_I  & =  x k_1^\mu + r_1 k_3^\mu  + r_2 k_4^\mu  + z k_2^\mu \, , \nonumber \\
   p_J  & =   (1-x) k_1^\mu + (1-r_1) k_3^\mu  + (1-r_2) k_4^\mu + (1-z) k_2^\mu \, ,
   \label{eq:42intma}
  \end{flalign}
 and the parameters $x,r_1,r_2,z$ are given\footnote{There are a few typos in the $4\to 2$ mapping formulas in Appendix B.1.1 of 
   \cite{Abelof:2011ap} as compared to the formulas of  \cite{GehrmannDeRidder:2007jk}.
The parameter $r_1$ should read $r_1 = (s_{jk} + s_{jl})/(s_{ij}+s_{jk}+s_{jl})$.
The term proportional to  $(r_1-r_2)$ in the definition of the parameter $z$ should read
$-(r_1 - r_2)(s_{ij} s_{kl} - s_{ik} s_{jl})/s_{il}.$} in appendix B.2.1 of  \cite{Abelof:2011ap}.
 For general configurations $k_j$ the mapped momenta are not on the $m_Q$ mass shell, $p^2_I, p_J^2 \neq m_Q^2$.
Four-momentum conservation in the   $e^-e^+$ c.m. frame reads
\begin{equation}\label{eq:4to2moco}
 \sqrt{s} = p_{0I} +  p_{0J} \, , \quad 0 = {\bf p}_{I} +  {\bf p}_{J} \, .
 \end{equation}
 Because of the second equation  the mapped momenta can be transformed without boost into on-shell four-momenta
 $k_{I}, k_{J}$ with mass $m_I = m_J = m_Q$. We have, analogous to the equations above,
   \begin{equation} \label{eq:scaletrnb} 
 {\bf k}_i = \xi {\bf p}_i \, , \quad k_{0i} = \sqrt{m_i^2 + \xi^2({p}_{0i}^2 - {p}^2_i)} \, , \qquad i = I,J \, ,
 \end{equation}
 where $\xi$ is the solution of the equation 
 \begin{equation} \label{eq:xisolnb} 
  \sqrt{s} = \sqrt{m_I^2 + \xi^2({p}_{0I}^2 - {p}^2_I)} + \sqrt{m_J^2 + \xi^2({p}_{0J}^2 - {p}^2_J)} \, .
  \end{equation}
 As in the case of equation \eqref{eq:xisol3p}  the iterative solution of  \eqref{eq:xisolnb} can be started with the
 value $\xi =\sqrt{(1-(2m_Q/\sqrt{s})^2)}.$ A few iterations $(n\leq 6)$ are enough 
 to get an accuracy of $10^{-14}\sqrt{s}/[{\rm GeV}]$. The mapped momenta $k_I, k_J$ converge to the correct momenta in the double unresolved limits. \\
 The mapped momenta required for $d \sigma^{S,b,2, Q \bar{Q} g g}_\mss{NNLO}$ of eq.~\eqref{sub:ggb2} are obtained in the same fashion. \\
 Abelof and Gehrmann-De Ridder have derived analytic formulas for the  on-shell massive 4-momenta $k^\mu_I, k^\mu_J$  \cite{AbelGerh2016}.
 
 Moreover, for the antenna subtraction term
 $d\sigma^{S,b,1, Q \bar{Q} q \bar{q}}_\mss{NNLO}$ of \eqref{eq:Sb1qqbar}
  two iterated $3\to 2$ momentum mappings are needed:
 \begin{equation}\label{eq:it1}
  1_Q, 3_q, 4_{\bar{q}} \: \text{with spectator} \:  2_{\bar{Q}} \stackrel{\rm I}{\longrightarrow}
  \widetilde{(13)}_Q, \widetilde{(34)}_g,~2_{\bar{Q}} \stackrel{\rm II}{\longrightarrow}
  \widetilde{(\widetilde{(13)}\widetilde{(34)})_Q},\widetilde{(\widetilde{(34)}2)_{\bar Q}} 
 \end{equation}
 and 
 \begin{equation}\label{eq:it2}
  2_{\bar{Q}}, 4_{\bar{q}}, 3_q  \:\text{with spectator} \:  1_{{Q}}  \stackrel{\rm I}{\longrightarrow}
   1_{{Q}},~\widetilde{(34)}_g, \widetilde{(42)}_{\bar{Q}}  \stackrel{\rm II}{\longrightarrow}
  \widetilde{(1\widetilde{(34)})_Q}, \widetilde{(\widetilde{(34)}\widetilde{(42)})_{\bar Q}}  \, . 
  \end{equation}
Let's consider \eqref{eq:it1}. The $3\to 2$ mapping I is done as described below eq.~\eqref{eq:32a1}: boost to the rest frame of
$Q_2^\mu$, rescale, and then boost back to the   $e^-e^+$ c.m. frame. The 
 rescaling involved in the subsequent mapping II is done directly in the 
 $e^-e^+$ c.m. frame. This yields the two mapped on-shell momenta with mass $m_Q$ on the right-hand side of \eqref{eq:it1}.
 The iterated $3\to 2$ mappings \eqref{eq:it2} and those involved in constructing the antenna subtraction term 
 $d\sigma^{S,b,1, Q \bar{Q} g g}_\mss{NNLO}$ of \eqref{sub:ggb1} are performed in analogous fashion. 
 
\acknowledgments
 We thank G. Abelof and A. Gehrmann-De Ridder for informing us about their phase-space mappings
  and for discussions. 
 L. Chen is supported by a scholarship  from the China Scholarship Council (CSC).
  D. Heisler is supported  by Deutsche Forschungsgemeinschaft through Graduiertenkolleg GRK 1675. 
  The work  of Z.G. Si  is supported by Natural Science Foundation of China (NSFC) and by Natural Science Foundation of
Shandong Province.




\begin{thebibliography}{99}





\bibitem{AguilarSaavedra:2001rg}
  J.~A.~Aguilar-Saavedra {\it et al.} [ECFA/DESY LC Physics Working Group Collaboration],
  \emph{TESLA: The Superconducting electron positron linear collider with an integrated x-ray laser laboratory.
  Technical design report. Part 3. Physics at an $e^+ e^-$ linear collider,}
  hep-ph/0106315.

\bibitem{Baer:2013cma}
  H.~Baer {\it et al.},
  \emph{The International Linear Collider Technical Design Report - Volume 2: Physics,}
  arXiv:1306.6352 [hep-ph].
 

\bibitem{Gomez-Ceballos:2013zzn}
  M.~Bicer {\it et al.} [TLEP Design Study Working Group Collaboration],
  JHEP {\bf 1401} (2014) 164
  doi:10.1007/JHEP01(2014)164
  [arXiv:1308.6176 [hep-ex]].
  
\bibitem{Moortgat-Picka:2015yla}
  G.~Moortgat-Pick {\it et al.},
  \emph{Physics at the $e^+ e^-$ Linear Collider,}
  Eur.\ Phys.\ J.\ C {\bf 75} (2015) no.8,  371
  doi:10.1140/epjc/s10052-015-3511-9
  [arXiv:1504.01726 [hep-ph]].
   
\bibitem{Vos:2016til}
  M.~Vos {\it et al.},
  \emph{Top physics at high-energy lepton colliders,}
  arXiv:1604.08122 [hep-ex].
  


\bibitem{Beneke:2015kwa}
  M.~Beneke, Y.~Kiyo, P.~Marquard, A.~Penin, J.~Piclum and M.~Steinhauser,
  \emph{Next-to-Next-to-Next-to-Leading Order QCD Prediction for the Top Antitop $S$-Wave 
   Pair Production Cross Section Near Threshold in $e^+e^-$ Annihilation,}
  Phys.\ Rev.\ Lett.\  {\bf 115} (2015) no.19,  192001
  doi:10.1103/PhysRevLett.115.192001
  [arXiv:1506.06864 [hep-ph]].




\bibitem{Jersak:1981sp}
  J.~Jersak, E.~Laermann and P.~M.~Zerwas,
  \emph{Electroweak Production of Heavy Quarks in $e^+ e^-$ Annihilation,}
  Phys.\ Rev.\ D {\bf 25} (1982) 1218
   Erratum: [Phys.\ Rev.\ D {\bf 36} (1987) 310].
  doi:10.1103/PhysRevD.36.310, 10.1103/PhysRevD.25.1218
  
\bibitem{Bernreuther:1997jn}
  W.~Bernreuther, A.~Brandenburg and P.~Uwer,
  \emph{Next-to-leading order QCD corrections to three jet cross-sections with massive quarks,}
  Phys.\ Rev.\ Lett.\  {\bf 79} (1997) 189
  doi:10.1103/PhysRevLett.79.189
  [hep-ph/9703305].

\bibitem{Brandenburg:1997pu}
  A.~Brandenburg and P.~Uwer,
  \emph{Next-to-leading order QCD corrections and massive quarks in $e^+ e^- \to$ three jets,}
  Nucl.\ Phys.\ B {\bf 515} (1998) 279
  doi:10.1016/S0550-3213(97)00790-6
  [hep-ph/9708350].
  
  
\bibitem{Rodrigo:1997gy}
  G.~Rodrigo, A.~Santamaria and M.~S.~Bilenky,
  \emph{Do the quark masses run? Extracting ${\bar m}_b(m_Z)$ from LEP data,}
  Phys.\ Rev.\ Lett.\  {\bf 79} (1997) 193
  doi:10.1103/PhysRevLett.79.193
  [hep-ph/9703358].
  
\bibitem{Rodrigo:1999qg}
  G.~Rodrigo, M.~S.~Bilenky and A.~Santamaria,
  \emph{Quark mass effects for jet production in $e^+ e^-$ collisions at the next-to-leading order: Results and applications,}
  Nucl.\ Phys.\ B {\bf 554} (1999) 257
  doi:10.1016/S0550-3213(99)00293-X
  [hep-ph/9905276].
  
\bibitem{Nason:1997tz}
  P.~Nason and C.~Oleari,
  \emph{Next-to-leading order corrections to momentum correlations in $Z^0 \to b{\bar b}$,}
  Phys.\ Lett.\ B {\bf 407} (1997) 57
  doi:10.1016/S0370-2693(97)00721-1
  [hep-ph/9705295].
  
\bibitem{Nason:1997nw}
  P.~Nason and C.~Oleari,
  \emph{Next-to-leading order corrections to the production of heavy flavor jets in $e^+ e^-$ collisions,}
  Nucl.\ Phys.\ B {\bf 521} (1998) 237
  doi:10.1016/S0550-3213(98)00125-4
  [hep-ph/9709360].
  
 
\bibitem{Beenakker:1991ca}
  W.~Beenakker, S.~C.~van der Marck and W.~Hollik,
  \emph{$e^+ e^-$ annihilation into heavy fermion pairs at high-energy colliders,}
  Nucl.\ Phys.\ B {\bf 365} (1991) 24.
  doi:10.1016/0550-3213(91)90606-X

\bibitem{Fleischer:2003kk}
  J.~Fleischer, A.~Leike, T.~Riemann and A.~Werthenbach,
  \emph{Electroweak one loop corrections for $e^+ e^-$ annihilation into t anti-top including hard bremsstrahlung,}
  Eur.\ Phys.\ J.\ C {\bf 31} (2003) 37
  doi:10.1140/epjc/s2003-01263-8
  [hep-ph/0302259].
  
\bibitem{Hahn:2003ab}
  T.~Hahn, W.~Hollik, A.~Lorca, T.~Riemann and A.~Werthenbach,
  \emph{$O(\alpha)$ electroweak corrections to the processes $e^+ e^- \to \tau^- \tau^+, c {\bar c}, b {\bar b}, t {\bar t}:$ A Comparison,}
  hep-ph/0307132.
  
  
\bibitem{Khiem:2012bp}
  P.~H.~Khiem {\it et al.},
  \emph{Full $\mathcal{O}(\alpha)$ electroweak radiative corrections to $e^+e^- \rightarrow t \bar{t} \gamma$ with GRACE-Loop,}
  Eur.\ Phys.\ J.\ C {\bf 73} (2013) no.4,  2400
  doi:10.1140/epjc/s10052-013-2400-3
  [arXiv:1211.1112 [hep-ph]].

  
\bibitem{Nejad:2016bci}
  B.~C.~Nejad, W.~Kilian, J.~M.~Lindert, S.~Pozzorini, J.~Reuter and C.~Weiss,
  \emph{NLO QCD Predictions for off-shell $t \bar t$ and $t \bar t H$ Production and Decay at a Linear Collider,}
  arXiv:1609.03390 [hep-ph].
  


  
\bibitem{Gorishnii:1986pz} 
  S.~G.~Gorishnii, A.~L.~Kataev and S.~A.~Larin,
  \emph{Three Loop Corrections of Order $M^2$ to the Correlator of
   Electromagnetic Quark Currents,}
  Nuovo Cim.\ A {\bf 92}  (1986) 119.
  
 
\bibitem{Chetyrkin:1996cf}
  K.~G.~Chetyrkin, J.~H.~K\"uhn and M.~Steinhauser,
  \emph{Three loop polarization function and  $O(\alpha_s^2)$ corrections to the production of heavy quarks,}
  Nucl.\ Phys.\ B {\bf 482} (1996) 213
  [hep-ph/9606230].


\bibitem{Chetyrkin:1997qi}
  K.~G.~Chetyrkin, R.~Harlander, J.~H.~K\"uhn and M.~Steinhauser,
  \emph{Mass corrections to the vector current correlator,}
  Nucl.\ Phys.\ B {\bf 503} (1997) 339
  [hep-ph/9704222].
  

\bibitem{Chetyrkin:1997pn}
  K.~G.~Chetyrkin, A.~H.~Hoang, J.~H.~K\"uhn, M.~Steinhauser and T.~Teubner,
  \emph{Massive quark production in electron positron annihilation to order $\alpha_s^2$,}
  Eur.\ Phys.\ J.\ C {\bf 2} (1998) 137
  [hep-ph/9711327].

\bibitem{Kiyo:2009gb}
  Y.~Kiyo, A.~Maier, P.~Maierhofer and P.~Marquard,
  \emph{Reconstruction of heavy quark current correlators at $O(\alpha_s^3)$,}
  Nucl.\ Phys.\ B {\bf 823} (2009) 269
  doi:10.1016/j.nuclphysb.2009.08.010
  [arXiv:0907.2120 [hep-ph]].
  
 
 
\bibitem{Dekkers:2014hna}
  O.~Dekkers and W.~Bernreuther,
  \emph{The real-virtual antenna functions for $S \to Q\bar{Q} X$ at NNLO QCD,}
  Phys.\ Lett.\ B {\bf 738} (2014) 325
  doi:10.1016/j.physletb.2014.09.060
  [arXiv:1409.3124 [hep-ph]].
 

 

\bibitem{Gao:2014nva}
  J.~Gao and H.~X.~Zhu,
  \emph{Electroweak prodution of top-quark pairs in $e^+e^-$ annihilation at NNLO in QCD: the vector contributions,}
  Phys.\ Rev.\ D {\bf 90} (2014) no.11,  114022
  doi:10.1103/PhysRevD.90.114022
  [arXiv:1408.5150 [hep-ph]].
  

\bibitem{Gao:2014eea}
  J.~Gao and H.~X.~Zhu,
  \emph{Top Quark Forward-Backward Asymmetry in $e^+e^-$ Annihilation at Next-to-Next-to-Leading Order in QCD,}
  Phys.\ Rev.\ Lett.\  {\bf 113} (2014) no.26,  262001
  doi:10.1103/PhysRevLett.113.262001
  [arXiv:1410.3165 [hep-ph]].
  


  

 
\bibitem{Kosower:1997zr}
  D.~A.~Kosower,
  \emph{Antenna factorization of gauge theory amplitudes,}
  Phys.\ Rev.\  {\bf D57 } (1998)  5410-5416.
  [hep-ph/9710213].


\bibitem{Kosower:2003bh}
  D.~A.~Kosower,
  \emph{Antenna factorization in strongly ordered limits,}
  Phys.\ Rev.\ D {\bf 71} (2005) 045016
  [hep-ph/0311272].

\bibitem{GehrmannDeRidder:2005cm}
  A.~Gehrmann-De Ridder, T.~Gehrmann and E.~W.~N.~Glover,
  \emph{Antenna subtraction at NNLO,}
  JHEP {\bf 0509} (2005) 056
  [hep-ph/0505111].


\bibitem{Currie:2013vh} 
  J.~Currie, E.~W.~N.~Glover and S.~Wells,
 \emph{Infrared Structure at NNLO Using Antenna Subtraction,}
  JHEP {\bf 1304}, 066 (2013)
  [arXiv:1301.4693 [hep-ph]].






\bibitem{GehrmannDeRidder:2009fz}
  A.~Gehrmann-De Ridder and M.~Ritzmann,
  \emph{NLO Antenna Subtraction with Massive Fermions,}
  JHEP {\bf 0907} (2009) 041
  [arXiv:0904.3297 [hep-ph]].


\bibitem{Abelof:2011jv}
  G.~Abelof and A.~Gehrmann-De Ridder,
  \emph{Antenna subtraction for the production of heavy particles at hadron colliders,}
  JHEP {\bf 1104} (2011) 063
  [arXiv:1102.2443 [hep-ph]].


\bibitem{Abelof:2011ap}
  G.~Abelof and A.~Gehrmann-De Ridder,
  \emph{Double real radiation corrections to $t\bar{t}$ production at the LHC: the all-fermion processes,}
  JHEP {\bf 1204} (2012) 076
  [arXiv:1112.4736 [hep-ph]].
  
 
\bibitem{Abelof:2012he}
  G.~Abelof, O.~Dekkers and A.~Gehrmann-De Ridder,
  \emph{Antenna subtraction with massive fermions at NNLO: Double real initial-final configurations,}
  JHEP {\bf 1212} (2012) 107
  [arXiv:1210.5059 [hep-ph]].
  
 
\bibitem{Abelof:2014fza}
  G.~Abelof, A.~Gehrmann-De Ridder, P.~Maierhofer and S.~Pozzorini,
  \emph{NNLO QCD subtraction for top-antitop production in the $ q\overline{q} $ channel,}
  JHEP {\bf 1408} (2014) 035
  doi:10.1007/JHEP08(2014)035
  [arXiv:1404.6493 [hep-ph]].

  
\bibitem{Abelof:2014jna}
  G.~Abelof and A.~Gehrmann-De Ridder,
  \emph{Light fermionic NNLO QCD corrections to top-antitop production in the quark-antiquark channel,}
  JHEP {\bf 1412} (2014) 076
  doi:10.1007/JHEP12(2014)076
  [arXiv:1409.3148 [hep-ph]].
 
  
  
\bibitem{Abelof:2015lna}
  G.~Abelof, A.~Gehrmann-De Ridder and I.~Majer,
  \emph{Top quark pair production at NNLO in the quark-antiquark channel,}
  JHEP {\bf 1512} (2015) 074
  doi:10.1007/JHEP12(2015)074
  [arXiv:1506.04037 [hep-ph]].
  
 

\bibitem{Bernreuther:2011jt}
  W.~Bernreuther, C.~Bogner and O.~Dekkers,
  \emph{The real radiation antenna function for $S \to Q {\bar Q} q {\bar q}$ at NNLO QCD,}
  JHEP {\bf 1106} (2011) 032
  doi:10.1007/JHEP06(2011)032
  [arXiv:1105.0530 [hep-ph]].
  
 
\bibitem{Bernreuther:2013uma}
  W.~Bernreuther, C.~Bogner and O.~Dekkers,
  \emph{The real radiation antenna functions for $S\rightarrow Q\bar{Q}gg$ at NNLO QCD,}
  JHEP {\bf 1310} (2013) 161
  doi:10.1007/JHEP10(2013)161
  [arXiv:1309.6887 [hep-ph]].
  
  

\bibitem{Czakon:2010td}
  M.~Czakon,
  \emph{A novel subtraction scheme for double-real radiation at NNLO,}
  Phys.\ Lett.\ B {\bf 693} (2010) 259
  [arXiv:1005.0274 [hep-ph]].

\bibitem{Czakon:2011ve}
  M.~Czakon,
   \emph{Double-real radiation in hadronic top quark pair production as a proof of a certain concept,}
  Nucl.\ Phys.\ B {\bf 849} (2011) 250
  [arXiv:1101.0642 [hep-ph]].


\bibitem{Czakon:2013goa}
  M.~Czakon, P.~Fiedler and A.~Mitov,
   \emph{The total top quark pair production cross-section at hadron colliders through $O(\alpha_s^4)$,}
  Phys.\ Rev.\ Lett.\  {\bf 110} (2013) 252004
  [arXiv:1303.6254 [hep-ph]].
  
\bibitem{Czakon:2015owf} 
  M.~Czakon, D.~Heymes and A.~Mitov,
  \emph{High-precision differential predictions for top-quark pairs at the LHC,}
  Phys.\ Rev.\ Lett.\  {\bf 116}, no. 8, 082003 (2016)
  doi:10.1103/PhysRevLett.116.082003
  [arXiv:1511.00549 [hep-ph]].
  
 
    


\bibitem{Gao:2012ja}
  J.~Gao, C.~S.~Li and H.~X.~Zhu,
  \emph{Top Quark Decay at Next-to-Next-to Leading Order in QCD,}
  Phys.\ Rev.\ Lett.\  {\bf 110} (2013) no.4,  042001
  doi:10.1103/PhysRevLett.110.042001
  [arXiv:1210.2808 [hep-ph]].


\bibitem{vonManteuffel:2014mva}
  A.~von Manteuffel, R.~M.~Schabinger and H.~X.~Zhu,
  \emph{The two-loop soft function for heavy quark pair production at future linear colliders,}
  Phys.\ Rev.\ D {\bf 92} (2015) no.4,  045034
  doi:10.1103/PhysRevD.92.045034
  [arXiv:1408.5134 [hep-ph]].
  
 
   \bibitem{AbelGerh2016}
     G.~Abelof and A.~Gehrmann-De Ridder, \emph{private communication.}
     

 \bibitem{Weinzierl:2006ij}
  S.~Weinzierl,
  \emph{NNLO corrections to 2-jet observables in electron-positron annihilation,}
  Phys.\ Rev.\ D {\bf 74} (2006) 014020
  doi:10.1103/PhysRevD.74.014020
  [hep-ph/0606008].
   
   
  
\bibitem{GehrmannDeRidder:2007jk}
  A.~Gehrmann-De Ridder, T.~Gehrmann, E.~W.~N.~Glover and G.~Heinrich,
  \emph{Infrared structure of $e^+ e^- \to$ 3 jets at NNLO,}
  JHEP {\bf 0711} (2007) 058
  [arXiv:0710.0346 [hep-ph]].
  
  
\bibitem{Glover:2010im}
  E.~W.~Nigel Glover and J.~Pires,
  \emph{Antenna subtraction for gluon scattering at NNLO,}
  JHEP {\bf 1006} (2010) 096
  doi:10.1007/JHEP06(2010)096
  [arXiv:1003.2824 [hep-ph]].


  
\bibitem{Hagiwara:1990dx}
  K.~Hagiwara, T.~Kuruma and Y.~Yamada,
  \emph{Three jet distributions from the one loop $Z g g$ vertex at $e^+ e^-$ colliders,}
  Nucl.\ Phys.\ B {\bf 358} (1991) 80.
  doi:10.1016/0550-3213(91)90532-3
  
  


\bibitem{Bernreuther:2004ih}
  W.~Bernreuther, R.~Bonciani, T.~Gehrmann, R.~Heinesch, T.~Leineweber, P.~Mastrolia and E.~Remiddi,
  \emph{Two-loop QCD corrections to the heavy quark form-factors: The Vector contributions,}
  Nucl.\ Phys.\ B {\bf 706} (2005) 245
  doi:10.1016/j.nuclphysb.2004.10.059
  [hep-ph/0406046].


\bibitem{Bernreuther:2004th}
  W.~Bernreuther, R.~Bonciani, T.~Gehrmann, R.~Heinesch, T.~Leineweber, P.~Mastrolia and E.~Remiddi,
  \emph{Two-loop QCD corrections to the heavy quark form-factors: Axial vector contributions,}
  Nucl.\ Phys.\ B {\bf 712} (2005) 229
  doi:10.1016/j.nuclphysb.2005.01.035
  [hep-ph/0412259].
  
\bibitem{Bernreuther:2005rw}
  W.~Bernreuther, R.~Bonciani, T.~Gehrmann, R.~Heinesch, T.~Leineweber and E.~Remiddi,
  \emph{Two-loop QCD corrections to the heavy quark form-factors: Anomaly contributions,}
  Nucl.\ Phys.\ B {\bf 723} (2005) 91
  doi:10.1016/j.nuclphysb.2005.06.025
  [hep-ph/0504190].
  
  
\bibitem{Gluza:2009yy}
  J.~Gluza, A.~Mitov, S.~Moch and T.~Riemann,
  \emph{The QCD form factor of heavy quarks at NNLO,}
  JHEP {\bf 0907} (2009) 001
  [arXiv:0905.1137 [hep-ph]].
 

  
\bibitem{Agashe:2014kda}
  K.~A.~Olive {\it et al.} [Particle Data Group Collaboration],
  \emph{Review of Particle Physics,}
  Chin.\ Phys.\ C {\bf 38} (2014) 090001.
  doi:10.1088/1674-1137/38/9/090001
  
  
\bibitem{Remiddi:1999ew}
  E.~Remiddi and J.~A.~M.~Vermaseren,
  \emph{Harmonic polylogarithms,}
  Int.\ J.\ Mod.\ Phys.\ A {\bf 15} (2000) 725
  doi:10.1142/S0217751X00000367
  [hep-ph/9905237].
  
  

 \bibitem{Gehrmann:2001pz}
  T.~Gehrmann and E.~Remiddi,
  \emph{Numerical evaluation of harmonic polylogarithms,}
  Comput.\ Phys.\ Commun.\  {\bf 141} (2001) 296
   [hep-ph/0107173].
  
\bibitem{Maitre:2005uu}
  D.~Maitre,
  \emph{HPL, a mathematica implementation of the harmonic polylogarithms,}
  Comput.\ Phys.\ Commun.\  {\bf 174} (2006) 222
  doi:10.1016/j.cpc.2005.10.008
  [hep-ph/0507152].

  



  

\bibitem{Blumlein:2009ta}
  J.~Bl\"umlein,
  \emph{Structural Relations of Harmonic Sums and Mellin Transforms up to Weight w = 5,}
  Comput.\ Phys.\ Commun.\  {\bf 180} (2009) 2218
  doi:10.1016/j.cpc.2009.07.004
  [arXiv:0901.3106 [hep-ph]].
  



\bibitem{Ablinger:2011te}
  J.~Ablinger, J.~Bl\"umlein and C.~Schneider,
  \emph{Harmonic Sums and Polylogarithms Generated by Cyclotomic Polynomials,}
  J.\ Math.\ Phys.\  {\bf 52} (2011) 102301
  doi:10.1063/1.3629472
  [arXiv:1105.6063 [math-ph]].
  
\bibitem{Ablinger:2013cf}
  J.~Ablinger, J.~Bl\"umlein and C.~Schneider,
  \emph{Analytic and Algorithmic Aspects of Generalized Harmonic Sums and Polylogarithms,}
  J.\ Math.\ Phys.\  {\bf 54} (2013) 082301
  doi:10.1063/1.4811117
  [arXiv:1302.0378 [math-ph]].




  
\bibitem{Czarnecki:1997vz}
  A.~Czarnecki and K.~Melnikov,
  \emph{Two loop QCD corrections to the heavy quark pair production cross-section
   in $e^+ e^-$ annihilation near the threshold,}
  Phys.\ Rev.\ Lett.\  {\bf 80} (1998) 2531
  [hep-ph/9712222].



\bibitem{Beneke:1997jm} 
  M.~Beneke, A.~Signer and V.~A.~Smirnov,
  \emph{Two loop correction to the leptonic decay of quarkonium,}
  Phys.\ Rev.\ Lett.\  {\bf 80}, 2535 (1998)
  [hep-ph/9712302].

\bibitem{Hoang:1997sj} 
  A.~H.~Hoang,
   \emph{Two loop corrections to the electromagnetic vertex for energies close
   to threshold},
  Phys.\ Rev.\ D {\bf 56} (1997) 7276 
  [hep-ph/9703404].
 
 
   
  
  
\bibitem{Bernreuther:2006vp}
  W.~Bernreuther, R.~Bonciani, T.~Gehrmann, R.~Heinesch, T.~Leineweber, P.~Mastrolia and E.~Remiddi,
  \emph{Two-Parton Contribution to the Heavy-Quark Forward-Backward Asymmetry in NNLO QCD,}
  Nucl.\ Phys.\ B {\bf 750} (2006) 83
  doi:10.1016/j.nuclphysb.2006.05.031
  [hep-ph/0604031].
  
%
 

\bibitem{Jersak:1979uv}
  J.~Jersak, E.~Laermann and P.~M.~Zerwas,
  \emph{QCD Corrected Forward Backward Asymmetry of Quark Jets in $e^+ e^-$ Annihilation,}
  Phys.\ Lett.\  {\bf 98B} (1981) 363.
  doi:10.1016/0370-2693(81)90926-6
  
\bibitem{Arbuzov:1991pr}
  A.~B.~Arbuzov, D.~Y.~Bardin and A.~Leike,
  \emph{Analytic final state corrections with cut for $e^+ e^- \to$ massive fermions,}
  Mod.\ Phys.\ Lett.\ A {\bf 7} (1992) 2029
   Erratum: [Mod.\ Phys.\ Lett.\ A {\bf 9} (1994) 1515].
  doi:10.1142/S0217732392001762
  
\bibitem{Djouadi:1994wt}
  A.~Djouadi, B.~Lampe and P.~M.~Zerwas,
  \emph{A Note on the QCD corrections to forward - backward asymmetries of heavy quark jets in Z decays,}
  Z.\ Phys.\ C {\bf 67} (1995) 123
  doi:10.1007/BF01564827
  [hep-ph/9411386].
  

  
\bibitem{Bardin:1999yd}
  D.~Y.~Bardin, P.~Christova, M.~Jack, L.~Kalinovskaya, A.~Olchevski, S.~Riemann and T.~Riemann,
  \emph{ZFITTER v.6.21: A Semianalytical program for fermion pair production in $e^+ e^-$ annihilation,}
  Comput.\ Phys.\ Commun.\  {\bf 133} (2001) 229
  doi:10.1016/S0010-4655(00)00152-1
  [hep-ph/9908433].
  
 
  
\bibitem{Altarelli:1992fs}
  G.~Altarelli and B.~Lampe,
  \emph{Second order QCD corrections to heavy quark forward - backward asymmetries,}
  Nucl.\ Phys.\ B {\bf 391} (1993) 3.
  doi:10.1016/0550-3213(93)90138-F
  
\bibitem{Ravindran:1998jw}
  V.~Ravindran and W.~L.~van Neerven,
  \emph{Second order QCD corrections to the forward - backward asymmetry in $e^+ e^-$ collisions,}
  Phys.\ Lett.\ B {\bf 445} (1998) 214
  doi:10.1016/S0370-2693(98)01436-1
  [hep-ph/9809411].
  
\bibitem{Catani:1999nf}
  S.~Catani and M.~H.~Seymour,
  \emph{Corrections of $O(\alpha_s^2)$ to the forward backward asymmetry,}
  JHEP {\bf 9907} (1999) 023
  doi:10.1088/1126-6708/1999/07/023
  [hep-ph/9905424].
  
  
\bibitem{Weinzierl:2006yt}
  S.~Weinzierl,
  \emph{The forward-backward asymmetry at NNLO revisited,}
  Phys.\ Lett.\ B {\bf 644} (2007) 331
  doi:10.1016/j.physletb.2006.11.076
  [hep-ph/0609021].

 
\bibitem{Waning2011}
K.~Waninger, \emph{Die Vorw\"arts-R\"uckw\"arts-Asymmetrie f\"ur schwere Quarks zur Ordnung $\alpha_s^2$},
doctoral thesis, RWTH Aachen University (2011).


\bibitem{Catani:2002hc}
  S.~Catani, S.~Dittmaier, M.~H.~Seymour and Z.~Trocsanyi,
  \emph{The Dipole formalism for next-to-leading order QCD calculations with massive partons,}
  Nucl.\ Phys.\ B {\bf 627} (2002) 189
  doi:10.1016/S0550-3213(02)00098-6
  [hep-ph/0201036].


  

     
\bibitem{Kleiss:1985gy}
  R.~Kleiss, W.~J.~Stirling and S.~D.~Ellis,
  \emph{A New Monte Carlo Treatment of Multiparticle Phase Space at High-energies,}
  Comput.\ Phys.\ Commun.\  {\bf 40} (1986) 359.
  doi:10.1016/0010-4655(86)90119-0

\bibitem{Devetak:2010na}
  E.~Devetak, A.~Nomerotski and M.~Peskin,
  \emph{Top quark anomalous couplings at the International Linear Collider,}
  Phys.\ Rev.\ D {\bf 84} (2011) 034029
  doi:10.1103/PhysRevD.84.034029
  [arXiv:1005.1756 [hep-ex]].
  
\bibitem{Rontsch:2015una}
  R.~R\"ontsch and M.~Schulze,
  \emph{Probing top-Z dipole moments at the LHC and ILC,}
  JHEP {\bf 1508} (2015) 044
  doi:10.1007/JHEP08(2015)044
  [arXiv:1501.05939 [hep-ph]].
  
  
\bibitem{Khiem:2015ofa}
  P.~H.~Khiem, E.~Kou, Y.~Kurihara and F.~Le Diberder,
  \emph{Probing New Physics using top quark polarization in the $e^+e^- \to t \bar{t}$ process at future Linear Colliders,}
  arXiv:1503.04247 [hep-ph].
  
\bibitem{Janot:2015yza}
  P.~Janot,
  \emph{Top-quark electroweak couplings at the FCC-ee,}
  JHEP {\bf 1504} (2015) 182
  doi:10.1007/JHEP04(2015)182
  [arXiv:1503.01325 [hep-ph]].
  
  
\bibitem{Amjad:2015mma}
  M.~S.~Amjad {\it et al.},
  \emph{A precise characterisation of the top quark electro-weak vertices at the ILC,}
  Eur.\ Phys.\ J.\ C {\bf 75} (2015) no.10,  512
  doi:10.1140/epjc/s10052-015-3746-5
  [arXiv:1505.06020 [hep-ex]].
  
 




 
\end{thebibliography}
\end{document}